\documentclass[fp,twocolumn]{jpsj3}
\usepackage{txfonts}
\usepackage{url}
\usepackage{xcolor}

\usepackage{bm}
\usepackage{amsfonts}
\usepackage{physics}
\usepackage{float}
\usepackage{here}
\usepackage{comment}
\usepackage{ulem}
\usepackage[whole]{bxcjkjatype}
\usepackage{bm}
\usepackage[ruled]{algorithm2e}
\usepackage{framed}

\usepackage{color}

\title{Effectiveness of Cardinality-Return Weighted Maximum Independent Set Approach for Financial Portfolio Optimization}

\author{Keita Takahashi$^{1}$, Tetsuro Abe$^{1}$, Yasuhito Nakamura$^{2}$, Ryo Hidaka$^{2}$, Shuta Kikuchi$^{1,3}$, and Shu Tanaka$^{1,3,4,5}$\thanks{shu.tanaka@keio.jp}}
\inst{
$^1$Graduate School of Science and Technology, Keio University, Kanagawa
223-8522, Japan \\
$^2$New Business Development Group, Data Business Promotion Department, ICT Solutions Division, Toshiba Digital Solutions Corporation, Kanagawa 212-8585, Japan\\
$^3$Department of Applied Physics and Physico-Informatics, Keio University, Kanagawa 223-8522, Japan \\
$^4$Keio University Sustainable Quantum Artificial Intelligence Center (KSQAIC), Keio University, Tokyo 108-8345,
Japan \\
$^5$Human Biology-Microbiome-Quantum Research Center (WPI-Bio2Q), Keio University, Tokyo 108-8345, Japan
}

\abst{
Markowitz's mean-variance model has fundamental limitations, including the assumption of normal return distributions and sensitivity to parameter estimation errors. To address these, we propose a novel graph-theoretic approach: the cardinality-return weighted maximum independent set (CR-WMIS) model. CR-WMIS integrates the risk diversification of the maximum independent set (MIS) model, selecting the largest number of weakly correlated stocks, with the expected return weighting of the weighted maximum independent set (WMIS) model. We validated the return and risk performance of this method through a five-year backtesting simulation (April 2019–March 2024) using S\&P 500 data and a simulated-bifurcation-based solver. Importantly, we conducted a comprehensive risk assessment, a facet insufficiently explored in previous studies. Results demonstrate that CR-WMIS exhibits superior return and risk characteristics compared to conventional MIS, WMIS, and the S\&P 500 index. This study provides a practical portfolio optimization method that overcomes the mean-variance model's theoretical limitations, contributing to both academic theory and investment decision-making.
}

\kword{Ising machine, Market graph, Maximum independent set, Portfolio optimization}

\begin{document}
\maketitle

\section{Introduction}
\label{sec: introduction}

The portfolio optimization problem, which involves selecting investment targets from a multitude of stocks in the market and allocating assets, has long been studied for both asset management firms and individual investors. A pioneering contribution in this field is modern portfolio theory (MPT), proposed by Markowitz~\cite{markowitz1952portfolio}. MPT provides a theoretical framework that aims not only to maximize expected returns but also to simultaneously minimize risk, based on the assumption that investors are risk-averse.

The mathematical model that embodies this theory is the mean-variance model. This model is formulated as a multi-objective optimization problem that maximizes the portfolio's expected return while minimizing the variance of returns (captured through the covariance between assets). However, the mean-variance model presents two significant practical limitations. First, it assumes that returns follow a normal distribution. In actual financial markets, return distributions do not necessarily conform to a normal distribution and often exhibit extreme price fluctuations, rendering covariance a potentially inadequate risk measure. Second, the model's optimal solution is highly sensitive to minor changes in the input parameters, namely expected returns and covariances. Since these parameters are estimated from historical data and do not perfectly predict the future, their estimation errors can severely compromise the stability of the optimal portfolio.

To overcome these limitations, various methods have been studied~\cite{rockafellar1999optimization, rockafellar2001conditional, black1991asset, chen2020generalized, gorissen2015practical, boginski2004network, boginski2005statistical, marzec2016portfolio, hidaka2023correlation}. For example, to tackle the non-normality of returns, conditional value-at-risk (CVaR) optimization was introduced; instead of minimizing variance, it focuses on minimizing the average loss in worst-case scenarios, directly managing the tail risk that variance fails to capture~\cite{rockafellar1999optimization, rockafellar2001conditional}. Additionally, to address the high sensitivity to input parameters, the Black--Litterman model was developed; it counters estimation errors by anchoring the portfolio to a stable market equilibrium and then blending in an investor's specific views~\cite{black1991asset}. Offering a fundamentally different paradigm to the models mentioned above, another promising approach is based on graph theory. Specifically, it involves reformulating the portfolio optimization problem as a maximum independent set (MIS) problem~\cite{boginski2004network, boginski2005statistical, marzec2016portfolio, hidaka2023correlation}. The MIS problem is a combinatorial optimization problem that seeks to find a set of vertices with the maximum cardinality in a graph, such that no two vertices in the set are connected by an edge (i.e., they are independent). In this approach, the market is mapped onto an undirected graph called a ``market graph''~\cite{butenko2003maximum, boginski2004network, boginski2005statistical}. In the market graph, each vertex represents an individual stock, and an edge signifies a strong correlation between stocks. By constructing edges only between pairs of highly correlated stocks, solving the MIS problem becomes equivalent to selecting the largest possible number of stocks that are weakly correlated with each other.

This MIS approach adheres to the risk-aversion concept of the mean-variance model and strongly promotes diversification by maximizing the number of selected stocks. This strategy of diversifying across numerous low-correlation stocks provides an effective countermeasure to the limitations of the mean-variance model, as it both enhances robustness against estimation errors in input parameters and does not rely on the assumption of the normal return distribution. For this reason, an approach based on the maximum clique problem, which is the dual problem to MIS, has been similarly investigated~\cite{kalyagin2014market, boginski2014networkbased}. Furthermore, it is worth noting that the correlation between the cardinality and the portfolio risk has also been studied in the context of investment concentration~\cite{tada2017random, shinzato2017replica}.

Indeed, the effectiveness of the MIS-based approach has been demonstrated in several studies. For instance, Hidaka et al. reported that their portfolio simulation in a large-scale market using this method achieved returns that outperformed market indices of TOPIX~\cite{hidaka2023correlation}. Furthermore, to enhance return characteristics, an extension to the weighted maximum independent set (WMIS) problem has also been investigated~\cite{marzec2016portfolio}. This involves assigning the expected return of each stock as a weight to its corresponding vertex in the market graph and then selecting an independent set that maximizes the sum of these weights. However, the WMIS approach has not been subjected to a multifaceted evaluation of both its return and risk aspects, and a comprehensive validation of its effectiveness remains insufficient. Moreover, portfolio selection based on existing WMIS models deviates from the fundamental risk-aversion strategy of the MIS model, which is risk diversification by maximizing the number of selected stocks, and thus may fail to leverage the key advantages of the MIS framework. Therefore, in this study, we propose a new model, the cardinality-return WMIS (CR-WMIS) model, which fuses the principles of the MIS and WMIS models. Specifically, the CR-WMIS model is formulated to explicitly incorporate both the risk diversification principle from MIS, achieved by maximizing the number of selected stocks, and the profitability-enhancement principle from WMIS, driven by maximizing expected returns. To validate our approach, we conduct a comprehensive quantitative evaluation of the model's performance from both return and risk perspectives.
However, solving such models presents a significant computational difficulty. The approaches treated in this study, including MIS, WMIS, and CR-WMIS, are all classified as NP-hard combinatorial optimization problems, making it difficult to find optimal solutions for large-scale instances. In recent years, solvers inspired by the Ising model of physics, known as Ising machines, have garnered significant attention~\cite{johnson2011quantum, goto2019combinatorial, goto2021high, matsubara2020digital, yoshimura2020cmos, honjo2021spin, mohseni2022ising, fixstars_amplify}. These machines are promising to find high-quality solutions for large-scale problems that were previously computationally intractable within a practical timeframe. Notably, finance presents numerous tasks well-suited to these physics-inspired computational approaches, with portfolio optimization via quantum annealing serving as a prominent example, thus motivating considerable research in this interdisciplinary field~\cite{Orus2019}. Building upon this context, our study utilizes one such Ising machine to evaluate the effectiveness of the proposed CR-WMIS model using large-scale real market data.

The remainder of this paper is organized as follows. In Section~\ref{sec: preliminary}, we review the theoretical foundations of this study, including the combinatorial optimization problems of MIS and WMIS, the mean-variance model, and annealing solvers. Section~\ref{sec: method} describes our methodology, detailing the market graph construction, the quadratic unconstrained binary optimization (QUBO) formulation, our proposed CR-WMIS model, and the backtesting simulation procedure. In Section~\ref{sec: experimental setup}, we present the specific experimental settings, including the dataset, validation period, and various model and solver parameters. Section~\ref{sec: results} reports the results of our five-year backtesting simulation using S\&P 500 data. In Section~\ref{sec: discussion}, we provide a detailed analysis of these results and investigate the mechanisms behind the performance improvements of the CR-WMIS model. Finally, Section~\ref{sec: conclusion} concludes the paper and outlines future work.

\section{Preliminary}
\label{sec: preliminary}

This section outlines the theoretical background that forms the foundation of the proposed CR-WMIS model. First, Section~\ref{subsec: MIS and WMIS model} explains the mathematical formulation of the maximum independent set (MIS) and weighted maximum independent set (WMIS) problems, which are the core combinatorial optimization problems in this study. Next, Section~\ref{subsec: mean variance model} reviews Markowitz's mean-variance model and summarizes its theoretical framework. Finally, Section~\ref{subsec: cop and annealing} describes the fundamental principles of the annealing solvers used in this study for solving NP-hard combinatorial optimization problems.

\subsection{MIS and WMIS Model}
\label{subsec: MIS and WMIS model}

This section describes the MIS and WMIS problems, which are the combinatorial optimization problems that underlie our proposed model.
First, let an undirected graph be denoted by $G=(V,E)$, where $V=\{v_1, v_2, \dots, v_N\}$ is a set of $N$ vertices and $E$ is a set of edges connecting them. A subset of vertices $S \subseteq V$ is called an independent set if for any two distinct vertices $u, v \in S$, no edge exists between them; that is, it satisfies the following condition:
\begin{equation}
    \forall u, v \in S, (u,v) \notin E.
    \label{eq: u,v}
\end{equation}
Based on this definition, the MIS problem is the task of finding an independent set $S$ with the maximum cardinality, that is, the largest number of vertices, in a graph $G$. It is formulated as follows~\cite{karp1985fast}:
\begin{align}
    \label{eq: mis maximize}
    \text{maximize} \quad & |S|,\\
    \label{eq: mis constraint}
    \text{subject to} \quad & (u,v) \notin E \quad (\forall u, v \in S).
\end{align}
Next, we define the WMIS problem. In this problem, we assign a weight $r_i$ to each vertex $i \in V$. The objective is to maximize the sum of the weights of the vertices contained in a set while satisfying the independent set constraint. The problem is formulated as follows~\cite{basagni2001finding}:
\begin{align}
    \text{maximize} \quad &\sum_{i \in S}r_i, \\
    \text{subject to} \quad &(u, v) \notin E \quad (\forall u, v \in S).
\end{align}

\subsection{Mean Variance Model}
\label{subsec: mean variance model}

Markowitz's mean-variance model was the first formal framework to quantitatively evaluate the trade-off between a portfolio's return and risk~\cite{markowitz1952portfolio}. In this model, we consider a portfolio composed of $N$ assets. Let $\bm{w} = (w_1, w_2, \cdots, w_N)$ be the vector of investment weights for each asset, $\bm{\mu} = (\mu_1, \mu_2, \cdots, \mu_N)$ be the vector of expected returns, and $\Sigma$ be the covariance matrix of the asset returns. The optimal portfolio can be obtained by solving the following problem:
\begin{align}
    \label{eq: obj of MV model}
    \text{maximize} \quad & \lambda \bm{w}^\mathrm{T} \bm{\mu} - \left( 1-\lambda \right) \bm{w}^\mathrm{T} \Sigma \bm{w}, \\
    \text{subject to} \quad & \sum_{i=1}^N w_i = 1, \quad w_i \geq 0.
\end{align}
In the objective function, the first term, $\bm{w}^\mathrm{T} \bm{\mu}$, represents the portfolio's total expected return, while the second term, $\bm{w}^\mathrm{T} \Sigma \bm{w}$, represents its total risk (i.e., the total variance of portfolio returns). The coefficient $\lambda \in [0, 1]$ is a parameter that signifies the investor's risk tolerance, serving to balance the two conflicting objectives of maximizing expected return and minimizing risk.

Furthermore, as indicated by Eq.~\eqref{eq: obj of MV model}, the mean-variance model is inherently a continuous optimization model that optimizes the investment weights $w_i$. In practical asset management, however, the problem is often subject to real-world constraints that are discrete in nature, such as cardinality constraints~\cite{parizy2022cardinality}, which limit the number of selected stocks, and round-lot constraints~\cite{lwin2014learning, chou2017portfolio}, which require investments in minimum transaction units. To incorporate such practical requirements, many recent studies have reformulated the mean-variance model as a combinatorial optimization problem. These approaches can be broadly classified into two categories. The first addresses constraints on investment quantities (such as round-lots) by reformulating the problem to solve for a discrete number of units for each asset, typically using integer variables~\cite{lwin2014learning, buonaiuto2023best}. The second approach, which is the focus of this paper, frames the problem as a discrete choice of whether to include an asset in the portfolio. This selection problem is naturally modeled using binary variables ($x_i \in \{0, 1\}$), where $x_i = 1$ if stock $i$ is selected and $x_i=0$ otherwise~\cite{lang2022strategic, marzec2016portfolio, phillipson2021portfolio, hidaka2023correlation}. This framework is particularly well-suited for incorporating cardinality constraints. The MIS and CR-WMIS strategies in our research belong to this second category of approaches.

\subsection{Combinatorial Optimization and Annealing Solvers}
\label{subsec: cop and annealing}

The MIS, WMIS, and the discretized versions of the mean-variance models discussed in the previous sections are all classified as NP-hard combinatorial optimization problems. This implies that as the problem size increases, the computational time required to find an optimal solution grows exponentially. Consequently, finding the optimal solution for large-scale instances in polynomial time is exceedingly difficult, necessitating the use of high-performance heuristic algorithms.

Recently, annealing algorithms inspired by physics have attracted significant attention as promising approaches to solving such combinatorial optimization problems. Specifically, active areas of research include classical simulated annealing~\cite{kirkpatrick1983optimization}, quantum annealing~\cite{kadowaki1998quantum, johnson2011quantum, tanaka2017quantum, yulianti2022implementation, chakrabarti2023quantum}, and quantum-inspired algorithms that emulate quantum effects on classical computers~\cite{santoro2002theory, goto2019combinatorial, goto2021high, okuyama2017ising}. The fundamental strategy of these methods is to map the target combinatorial optimization problem onto a physical system, such as an Ising model or a quadratic unconstrained binary optimization (QUBO) formulation, and then search for the solution that corresponds to the lowest state of its energy function~\cite{lucas2014ising, tanahashi2019application, yarkoni2022quantum}. In recent years, Ising machines and related annealing-based solvers have been applied to a wide range of practical problems, including logistics~\cite{feld2019hybrid, kanai2024annealing}, materials science~\cite{harris2018phase, king2018observation, endo2022phase, sampei2023quantum, nawa2023quantum, honda2024development, xu2025quantum}, machine learning~\cite{nath2021review, liu2024implementation}, and black-box optimization~\cite{kitai2020designing, tamura2025black}, demonstrating their versatility beyond purely theoretical optimization tasks. In particular, when a problem is input in the QUBO format, which is highly relevant to this study, we work with an energy function of the following form:
\begin{equation}
    \label{eq: general qubo}
    H = \sum_{1 \leq i < j \leq N} a_{i,j} x_i x_j +  \sum_{i=1}^N b_i x_i.
\end{equation}
Here, $x_i \in \{0, 1\}$ are binary variables, $a_{i,j}$ represents the interaction strength between the variable pair $x_i$ and $x_j$, and $b_i$ is the coefficient for the linear term of each variable. By mapping the optimal solution of a combinatorial optimization problem to the solution that minimizes the function in Eq.~\eqref{eq: general qubo}, the search for the optimal solution is transformed into an energy minimization problem for $H$. Then, during the annealing process, the system explores optimal or near-optimal solutions by initially allowing for large fluctuations in the search and gradually reducing them over time.

\section{Method}
\label{sec: method}

This section details our proposed portfolio optimization method and the specific methodology used to validate its effectiveness. First, Section~\ref{subsec: market graph construction and QUBO formulation} explains the method for constructing the market graph, which is fundamental to our approach, and presents the quadratic unconstrained binary optimization (QUBO) formulations for the MIS, WMIS, and our proposed CR-WMIS models. Next, Section~\ref{subsec: backtest simulation methodology} describes the detailed procedures and settings for the backtesting simulation used to evaluate the models' performance using historical stock price data. Finally, Section~\ref{subsec: evaluation metrics} defines the various metrics used to perform a multifaceted evaluation of the simulation results from both return and risk perspectives.

\subsection{Market Graph Construction and QUBO Formulation}
\label{subsec: market graph construction and QUBO formulation}

In portfolio optimization based on the MIS and WMIS approaches, the financial market is mapped onto a market graph that represents the relationships among stocks. This section describes the detailed procedure for constructing this graph and presents the QUBO formulations for the MIS and WMIS models.

\subsubsection{Graph Construction}

The construction of the market graph begins with the calculation of returns, which quantifies the price changes of each stock. The return calculation interval can be chosen arbitrarily, and in this study, we define this period as $D_\mathrm{R}$ days. The $D_\mathrm{R}$-day return for stock $i$ on a given date $d$, denoted as $R_i(d)$, is defined using the following logarithmic return:
\begin{equation}
    R_i (d) = \ln{\frac{P_i (d)}{P_i (d - D_\mathrm{R})}}.
    \label{eq: log return}
\end{equation}
Here, $P_i(d)$ is the price of stock $i$ on date $d$. Logarithmic returns are widely used in financial engineering due to their mathematical convenience, such as allowing multi-period returns to be handled through simple addition~\cite{miskolczi2017note}.
In this paper, we use the uppercase letter $D$ to denote a duration in days and the lowercase letter $d$ to denote a specific date. Consequently, the date that is $D$ days prior to $d$ is expressed as $d-D$. It should be noted that all durations and dates mentioned in this paper refer exclusively to trading days.

Next, we compute the pairwise correlation coefficients between stocks using the calculated return time-series data. For each pair of stocks $i$ and $j$, the Pearson correlation coefficient $C_{i,j}$ is computed from the return data over the past $D_\mathrm{opt}$ trading days:
\begin{equation}
  C_{i,j} =
  \frac{
    \displaystyle \sum_{d=d_\mathrm{s} + D_\mathrm{R}}^{d_\mathrm{o}} 
    \left( R_i (d) - \bar{R}_i \right) \left( R_j (d) - \bar{R}_j \right)
  }
  {
    \sqrt{
      \displaystyle \sum_{d=d_\mathrm{s} + D_\mathrm{R}}^{d_\mathrm{o}} 
      \left( R_i (d) - \bar{R}_i \right)^2
    }
    \sqrt{
      \displaystyle \sum_{d=d_\mathrm{s} + D_\mathrm{R}}^{d_\mathrm{o}}
      \left( R_j (d) - \bar{R}_j \right)^2
    }
  }.
\end{equation}
Here, $d_\mathrm{s}$ is the start date for the calculation, and $d_\mathrm{o}$ is the optimization date, with the relationship $D_\mathrm{opt} = d_\mathrm{o} - d_\mathrm{s}$. Furthermore, $\bar{R}_i$ denotes the mean return of stock $i$, calculated as:
\begin{equation}
    \label{eq: simple average return}
    \bar{R}_i = \frac{1}{D_\mathrm{opt} - D_\mathrm{R}} \sum_{d=d_\mathrm{s} + D_\mathrm{R}}^{d_\mathrm{o}} R_i (d).
\end{equation}
The continuous correlation coefficients $C_{i,j}$ are subsequently binarized using a threshold $\theta$~\cite{boginski2004network, boginski2005statistical, marzec2016portfolio, hidaka2023correlation}. This step defines the elements $f_{i,j}$ of the graph's adjacency matrix as follows:
\begin{equation}
    f_{i,j} = \begin{cases}
        1 & \:\: \left( C_{i,j} \geq \theta \right) \\ 0 & \:\: \left( C_{i,j} < \theta \right).
    \end{cases}
\end{equation}
This operation constructs the market graph $G=(V,E)$. In other words, each stock $i$ is treated as a vertex $v_i \in V$, and an edge $(i,j) \in E$ is added between a pair of stocks $(i,j)$ only if their correlation is larger than or equal to the threshold $\theta$. By binarizing the correlation strength in this manner, we aim to explicitly cluster stocks with similar price movements, thereby enhancing the portfolio's diversification effect.

\subsubsection{QUBO Formulation of MIS and WMIS}

Based on the constructed market graph, and the objective functions and constraints introduced in Section~\ref{sec: preliminary}, we formulate the MIS and WMIS models in the QUBO format. First, we introduce a binary variable $x_i$, defined as $x_i=1$ if stock $i$ is included in the portfolio and $x_i=0$ otherwise. The MIS problem can then be expressed as the following QUBO-based minimization problem~\cite{lucas2014ising, hidaka2023correlation}:
\begin{equation}
\label{eq: mis qubo}
    H_\mathrm{MIS} = A \sum_{1 \leq i < j \leq N} f_{i,j} x_i x_j - B \sum_{i=1}^N x_i.
\end{equation}
Here, $N$ is the total number of stocks, and $A$ and $B$ are positive constants. The first term in this QUBO represents a penalty term that enforces the independent set constraint, corresponding to \eqref{eq: mis constraint}. The second term, on the other hand, is an objective term that encourages the maximization of the number of selected stocks, corresponding to \eqref{eq: mis maximize}. The balance between these two effects can be controlled by adjusting the ratio of the coefficients $A$ and $B$. Generally, $A$ must be set sufficiently larger than $B$ to ensure the independent set constraint is strictly satisfied.
Next, to improve profitability, we consider the WMIS model, which incorporates the expected return of each stock into the MIS model. The QUBO formulation for WMIS can be expressed as:
\begin{equation}
\label{eq: wmis qubo}
    H_\mathrm{WMIS} = A \sum_{1 \leq i < j \leq N} f_{i,j} x_i x_j - \mu_\mathrm{R} \sum_{i=1}^N r_i x_i.
\end{equation}
Here, $\mu_\mathrm{R}$ is the coefficient for the second term, and $r_i$ represents the expected return for stock $i$. The specific estimation methods for this quantity $r_i$ will be detailed in Section~\ref{subsubsec: expected return estimation}.  In this WMIS model, the first term acts as a risk-aversion component by enforcing the independent set constraint, while the second term aims to maximize the expected return. This formulation is very similar to models found in prior WMIS research~\cite{marzec2016portfolio} and to the maximum weighted independent set (MWIS) problem in the context of graph theory and optimization~\cite{basagni2001finding, choi2010adiabatic, hattori2025controlled}. As the first term promotes risk aversion and the second promotes return maximization, this model can be interpreted as a graph-theoretic representation of the mean-variance model.

\subsubsection{Proposed Model: Cardinality-Return WMIS (CR-WMIS)}

While the WMIS model seeks to enhance expected returns, its application to portfolio optimization has a significant drawback: it lacks the concept of ``diversification by maximizing the number of selected stocks'', which was the original motivation for introducing the MIS approach to the portfolio optimization problem. Since the WMIS model solely maximizes the sum of expected returns, it tends to yield solutions where investment is concentrated in a few stocks with extremely high expected returns $r_i$, even while satisfying the independent set constraint. This may undermine the inherent advantages of the MIS model, such as its robustness to non-normal return distributions and estimation errors in input parameters.

Therefore, to leverage both the risk-aversion property of MIS and the return-enhancement effect of WMIS, we propose the following new model:
\begin{align*}
\label{eq: cr-wmis qubo}
    H_\mathrm{CR-WMIS} &= A \sum_{1 \leq i < j \leq N} f_{i,j} x_i x_j - B \sum_{i=1}^N x_i
    - \mu_\mathrm{R} \sum_{i=1}^N r_i x_i \\
    &= A \sum_{1 \leq i < j \leq N} f_{i,j} x_i x_j - \sum_{i=1}^N (B + \mu_\mathrm{R} r_i) x_i.
        \stepcounter{equation}\tag{\theequation} 
\end{align*}
This model integrates the formulations of the MIS model expressed by Eq.~\eqref{eq: mis qubo} and the WMIS model expressed by Eq.~\eqref{eq: wmis qubo}, and we refer to it as the cardinality-return WMIS (CR-WMIS) model. The CR-WMIS model can be viewed either as the MIS model with an added weighting based on expected returns or as the WMIS model with a uniform offset introduced to maximize cardinality.
Thus, the CR-WMIS model explicitly incorporates the principles of both cardinality maximization and expected return maximization. It is designed to retain the robust risk diversification capability inherited from MIS while incorporating the potential for enhanced profitability from WMIS. Crucially, this model assumes that the coefficient $A$ is set sufficiently large relative to the other coefficients (i.e., $A \gg B$ and $A \gg \mu_\mathrm{R} r_i$) to ensure the independent set constraint is satisfied. Under this condition, the relative magnitudes between the coefficients $B$ and $\mu_\mathrm{R} r_i$ determines which objective is prioritized. When $B \gg \mu_\mathrm{R} r_i$, the model approximates the MIS model, whereas when $B \ll \mu_\mathrm{R} r_i$ it behaves similarly to the WMIS model.

\subsubsection{Expected Return Estimation}
\label{subsubsec: expected return estimation}

The accuracy of the estimation of the expected return $r_i$ in the existing WMIS and our proposed CR-WMIS models directly impacts portfolio performance. In this study, we consider two approaches for estimating this quantity.

The first approach is the simple average (SAvg), which is one of the most straightforward methods~\cite{welch2008comprehensive, ferreira2011forecasting}. This method uses the equally weighted mean of historical returns, which is equivalent to the mean return $\bar{R}_i$ already defined in \eqref{eq: simple average return}:
\begin{equation}
    \label{eq: sma}
    r_i^\mathrm{SAvg} = \bar{R}_i
\end{equation}
Here, $\bar{R}_i$ is calculated using the historical data over the $D_\mathrm{opt}$ days prior to the optimization date $d_\mathrm{o}$, as shown in \eqref{eq: simple average return}. Because this method assigns equal importance to all historical data points, it is less sensitive to noise and can stably capture long-term trends. However, it has the drawback of responding slowly to react to abrupt changes in the market environment.

The second approach is the exponentially weighted average (EWAvg) method. Similar to the exponentially weighted moving average (EWMA) used in volatility forecasting~\cite{jpmorgan1996riskmetrics}, this technique assigns exponentially decaying weights, thereby giving larger influence to more recent data. Our EWAvg approach is applied here to expected return estimation to capture the ``momentum effect'', the phenomenon where recent returns tend to predict future returns~\cite{moskowitz2012time, baltas2013momentum}. It can be calculated as follows:
\begin{equation}
    r_i^\mathrm{EWAvg} = \frac{\sum_{d=d_\mathrm{s} + D_\mathrm{R}}^{d_\mathrm{o}} (1-\alpha)^{d_\mathrm{o}-d} R_i(d)}{\sum_{d=d_\mathrm{s} + D_\mathrm{R}}^{d_\mathrm{o}} (1-\alpha)^{d_\mathrm{o}-d}}.
\end{equation}
The parameter $\alpha \in (0, 1)$ represents the decay parameter. A larger value of $\alpha$ increases the influence of recent data, while a smaller value allows the impact of past data to persist longer. Although the choice of $\alpha$ can be arbitrary, a common relationship exists between $\alpha$ and the length of the time series $D$ \cite{pandas-ewm-doc}:
\begin{equation}
    \label{eq: alpha eqma}
    \alpha = \frac{2}{D_\mathrm{opt}-D_\mathrm{R}+1}.
\end{equation}
This formula is designed such that the center of mass of the EWAvg's exponentially decaying weights coincides with that of the SAvg's uniform weights over the same period, $D_\mathrm{opt} - D_\mathrm{R}$. It is important to note, however, that because the weighting methods of SAvg and EWAvg are fundamentally different, the resulting values are not equivalent.

The selection of a method like SAvg or EWAvg requires careful consideration, since the temporal weighting used in estimating a portfolio's expected return should be determined appropriately based on factors such as market liquidity and the lookback period used for the estimation.

\subsection{Backtest Simulation Methodology}
\label{subsec: backtest simulation methodology}

The optimization models formulated in the previous section determine the optimal combination of stocks at a single point in time. However, to assess the practical effectiveness of our proposed method, it is necessary to validate its performance over multiple consecutive periods. Therefore, in this study, we perform backtesting using historical stock price data.

The detailed procedure of the simulation is outlined in Algorithm~\ref{algo: portfolio backtest}. The process is repeated on a monthly basis: at the end of each month, the constituent stocks of the portfolio are optimized and rebalanced. This newly constructed portfolio is then held throughout the following month. Transaction fees are incorporated into the analysis at the time of rebalancing.

\begin{algorithm}[t]
\SetAlgoLined
\KwData{Historical stock price data $P_i(d)$, where $P_i(d)$ denotes the closing price of stock $i$ on trading date $d$; initial capital $F_1$; backtesting horizon $M$; model parameters $D_{\mathrm{opt}}$, $\theta$}

\KwResult{Final capital $F_M$ and performance metrics}

\SetKwInOut{Input}{Input}
\SetKwInOut{Output}{Output}

Initialize month index $m = 1$\;

\While{$m < M$}{
    \textbf{QUBO Formulation:}
    Set the optimization date as $d_{m, \mathrm{final}}$, where $d_{m, \mathrm{final}}$ is the final trading day in month $m$)\;
    Calculate correlation matrix $C_{i,j}$ using past $D_{\mathrm{opt}}$ days of historical data\;
    Construct market graph with adjacency matrix $f_{i,j}$ using threshold $\theta$\;
    Estimate expected returns $r_i$ using either EWAvg or SAvg (Note: this step is only required for the WMIS and CR-WMIS models, not for the MIS model)\;
    Formulate QUBO\;
    
    \textbf{Portfolio Optimization:}
    Solve the optimization problem to obtain optimal portfolio $\bm{x}^* = (x_1^*, x_2^*, \ldots, x_N^*)$\;
    Define the set of selected stocks $S_{\mathrm{selected}} = \{i \mid x_i^* = 1\}$\;
    
    \textbf{Weight Allocation and Rebalancing:}
    Determine the capital allocation weights $\{w_p\}_{p \in S_{\mathrm{selected}}}$ such that $\sum_{p \in S_{\mathrm{selected}}} w_p = 1$\;
    
    \textbf{Portfolio Update:}
    Update capital;
    $F_{m+1} = F_m \times \left(\sum_{p \in S_{\mathrm{selected}}} w_p \times \frac{P_{p} (d_{m+1, \mathrm{final}})}{P_{p} (d_{m, \mathrm{final}})}\right)$\;
    
    \textbf{Increment time:}
    $m \leftarrow m + 1$;
}

\textbf{Return:} Final capital $F_M$ and corresponding backtesting results

\caption{Portfolio Backtesting Simulation Algorithm}
\label{algo: portfolio backtest}
\end{algorithm}

Furthermore, to assess the robustness of our method, we perform backtesting over multiple time windows using a rolling-window analysis~\cite{zivot2003rolling, amini2024investigating}.

It should be noted that the optimization models in this study determine only the set of stocks to be included in the portfolio. Therefore, the allocation of capital among the selected stocks must be decided separately. To investigate how the asset allocation strategy affects the overall portfolio performance, we examine two distinct strategies with different characteristics.

The first is the equal-weight (EW) strategy, which is one of the simplest and most widely used methods~\cite{demiguel2009optimal, hidaka2023correlation}. In this approach, given the solution $\bm{x}^* = (x_1^*, x_2^*, \ldots, x_N^*)$ obtained from the optimization, we define the set of selected stocks as $S_{\mathrm{selected}} = \{i \mid x_i^*=1\}$ and its cardinality as $N_{\mathrm{portfolio}} = |S_{\mathrm{selected}}|$. The weight $w_p^\mathrm{EW}$ for each stock $p \in S_{\mathrm{selected}}$ is determined as follows:
\begin{equation}
    w_p^\mathrm{EW} = \frac{1}{N_\mathrm{portfolio}}, \quad (p \in S_{\mathrm{selected}}).
\end{equation}
In this way, the EW strategy allocates capital equally across all selected stocks.

The second is a strategy that allocates capital based on the risk characteristics of each stock. Specifically, we employ the inverse-volatility-weight (IVW) strategy, which uses volatility as a risk measure~\cite{shimizu2020constructing, hidaka2023correlation}. This strategy aims to reduce overall portfolio risk by allocating more capital to low-volatility stocks and less to high-volatility stocks. 
As a risk metric, a lower volatility value indicates lower risk. 
Mathematically, volatility is calculated as the standard deviation of returns. Let $\sigma_i$ denote the volatility of stock $i$, it is computed as:
\begin{equation}
    \label{eq: historical volatility}
    \sigma_i = \sqrt{\frac{1}{D_\mathrm{opt}-D_\mathrm{R}}\sum_{d = d_\mathrm{s} + D_\mathrm{R}}^{d_\mathrm{o}}(R_i(d) - \bar{R}_i)^{2}}.
\end{equation}
The weight $w_p^\mathrm{IVW}$ for each stock $p \in S_{\mathrm{selected}}$ is then determined as:
\begin{equation}
    w_p^\mathrm{IVW} = \frac{\sigma_p^{-1}}{\sum_{l \in S_{\mathrm{selected}}} \sigma_l^{-1}}, \quad (p \in S_{\mathrm{selected}}).
\end{equation}
The denominator, representing the sum of the inverse volatilities for all selected stocks, serves to normalize the weights so that  the total investment sums to $1$.

\subsection{Evaluation Metrics}
\label{subsec: evaluation metrics}

This section describes the metrics used to evaluate the effectiveness of the proposed method from the multifaceted perspectives, including both return and risk.

\subsubsection{Return Metric (Cumulative Return)}

The cumulative return is the most fundamental profitability metric, representing the total rate of return on an investment over the entire investment period. The cumulative return at month $m$, denoted by $CR_m$, is calculated from the capital $F_m$ at month $m$ as follows:
\begin{equation}
    CR_m = \frac{F_m - F_1}{F_1}.
    \label{eq: cumret}
\end{equation}

\subsubsection{Risk Metrics}

\paragraph{Maximum Drawdown (MDD)}

The maximum drawdown (MDD) is a metric that indicates the largest decline from a peak to a trough in capital during an investment period~\cite{grossman1993optimal, yang2012optimal, drenovak2022mean}. It represents the maximum loss an investor could have experienced and is therefore a crucial indicator for assessing a portfolio's potential downside risk. The MDD can be calculated using the following formula:
\begin{equation}
    \mathrm{MDD} = - \underset{m \in [m_\mathrm{start}, m_\mathrm{end}]}{\mathrm{max}} \frac{\underset{\tau \in [m_\mathrm{start}, m]}{\mathrm{max} F_\tau} - F_m}{\underset{\tau \in [m_\mathrm{start}, m]}{\mathrm{max} F_\tau}}.
    \label{eq: mdd}
\end{equation}
Here, $m_{\mathrm{start}}$ and $m_{\mathrm{end}}$ denote the first and last months of the backtesting period, respectively. As MDD captures the magnitude of consecutive losses, that is, the downside deviation of returns, over a period, it is widely used as a standard risk metric.

\paragraph{Volatility}

As already mentioned in \eqref{eq: historical volatility}, volatility is defined as the standard deviation of returns and is quite a common measure of risk~\cite{moreira2017volatility, shimizu2020constructing}. Volatility of created portfolio can be computed as:
\begin{equation}
    \sigma= \sqrt{\frac{1}{D_\mathrm{eval}}\sum_{d=d_\mathrm{o}+1}^{d_\mathrm{o} + D_\mathrm{eval}}(R_{\mathrm{portfolio}}(d) - \bar{R}_\mathrm{portfolio})^2}.
\end{equation}
Here, $R_\mathrm{portfolio}(d)$ denotes the portfolio's return on trading day $d$, $\bar{R}_\mathrm{portfolio}$ its mean return, and $D_\mathrm{eval}$ the length of the evaluation period, which corresponds to the immediate future data following the optimization date $d_\mathrm{o}$. $R_\mathrm{portfolio}(d)$ and $\bar{R}_\mathrm{portfolio}$ are calculated as:
\begin{align}
    \label{eq: portfolio daily return}
    R_{\mathrm{portfolio}}(d) &= \sum_{p \in S_{\mathrm{selected}}} w_p R_p(d),\\
    \label{eq: portfolio mean return}
    \bar{R}_\mathrm{portfolio} &= \frac{1}{D_\mathrm{eval}} \sum_{d=d_\mathrm{o}+1}^{d_\mathrm{o}+D_\mathrm{eval}} R_{\mathrm{portfolio}}(d).
\end{align}
As a risk metric, a lower volatility value is considered more favorable. Although volatility is simple to calculate and interpret, it treats upward and downward movements equally, potentially overestimating the downside risk that truly concerns investors. Furthermore, because it assumes a normal distribution, volatility may fail to adequately capture extreme values or asymmetries.

\paragraph{Value at Risk (VaR)}
Value at Risk (VaR) represents the maximum potential loss that may be incurred over a given period at a specified confidence level $(1-q)$~\cite{jpmorgan1996riskmetrics}:
\begin{equation}
    P (R_\mathrm{portfolio} < \mathrm{VaR}_q) = q.
\end{equation}
Here, $P(\cdot)$ denotes the probability of an event. Although VaR is intuitive and easy to interpret, it has a significant limitation: it provides no information regarding the magnitude of losses exceeding the VaR threshold.

\paragraph{Conditional Value at Risk (CVaR)}

Conditional Value at Risk (CVaR), also known as the expected shortfall, is defined as the conditional expectation of losses exceeding the VaR level~\cite{rockafellar1999optimization, rockafellar2001conditional}:
\begin{equation}
    \mathrm{CVaR}_q = \mathrm{E}[R_\mathrm{portfolio} | R_\mathrm{portfolio} < \mathrm{VaR}_q].
\end{equation}
Here, E$[\cdot]$ denotes the expectation operator with respect to the distribution of portfolio returns. Because CVaR represents the average loss beyond the VaR threshold, it can quantify the magnitude of the downside risk that is overlooked by VaR.

\section{Experimental Setup}
\label{sec: experimental setup}

This section describes the settings for the backtesting simulation conducted to validate the effectiveness of our proposed method using real market data. First, Section~\ref{subsec: dataset and experimental period} defines the dataset and experimental period for our analysis. Next, Section~\ref{subsec: backtest protocol} describes the operational rules for the backtest, including the rebalancing strategy and transaction fees. Finally, Section~\ref{subsec: model and solver parameters} presents the values of the parameters used for model construction and optimization.

\subsection{Dataset and Experimental Period}
\label{subsec: dataset and experimental period}

To evaluate the performance of our proposed method in a real-world market, this study uses the constituent stocks of the S\&P 500, a representative index of the U.S. stock market. The validation period covers five years, from April 2019 to March 2024. This period encompasses a particularly diverse and complex range of market conditions. For example, it includes the sharp bear market from February to April 2020 triggered by the COVID-19 pandemic, the subsequent post-pandemic recovery, and the prolonged bear market from January to October 2022, driven by inflation and rising interest rates. Accordingly, our validation period contains both phases of abrupt market decline and sustained growth, making it a suitable timeframe for evaluating the robustness of portfolio optimization methods. To benchmark the performance of the CR-WMIS method, we compare it with the MIS and WMIS models, as well as the S\&P 500 index as a market benchmark.

\subsection{Backtest Protocol}
\label{subsec: backtest protocol}

The operational rules of our simulation were designed to reflect practical trading scenarios, with reference to prior study~\cite{hidaka2023correlation}. The basic strategy involves optimizing and rebalancing the portfolio at the end of each month. A uniform transaction fee of $0.1\%$ is applied to all trades executed during rebalancing. Furthermore, to evaluate the robustness of our method, we perform a rolling-window analysis, repeating the simulation for 2-year periods, with the start date of each period advanced by six months. In accordance with the monthly rebalancing framework, the performance of each portfolio constructed on a rebalancing date is evaluated using market data from the subsequent month ($D_\mathrm{eval}=20$ trading days). Note that for the calculation of risk metrics such as volatility, VaR, and CVaR, we use daily returns ($D_\mathrm{R} = 1$) to ensure a sufficient sample size. The detailed simulation parameters are summarized in Table~\ref{table: simulation parameters}.

\begin{table}[t]
\caption{Parameters used in the backtesting simulation}
\label{table: simulation parameters}
\setlength{\tabcolsep}{3pt}
\begin{tabular}{|p{80pt}p{150pt}|}
\hline
Rebalancing frequency & Monthly at month-end \\
Transaction fee & 0.1\% \\
Rolling-window & 2-year window, stepped every 6 months \\
Evaluation period $D_\mathrm{eval}$ & 1 month (20 trading days) post-optimization\\
Number of backtests & 10\\
\hline
\end{tabular}
\end{table}

\subsection{Model Parameters}
\label{subsec: model and solver parameters}

The parameters used in model construction are listed in Table~\ref{table: model parameters}.

\begin{table}[t]
\caption{Parameters used in model construction}
\label{table: model parameters}
\setlength{\tabcolsep}{3pt}
\begin{tabular}{|p{30pt}|p{200pt}|}
\hline
$A$ & 2 \\
$B$ & 1 \\
$\mu_\mathrm{R}$ & 2 \\
$\theta$ & 0.23 \\
$D_{\mathrm{R}}$ & 20 \\
$D_\mathrm{opt}$ & 3 years (756 trading days) prior to the optimization date\\
$\alpha$ & 0.003\\
\hline
\end{tabular}
\end{table}

The QUBO coefficients $A$ and $B$ are set such that $A > B$ to ensure that the weight of the independent set constraint term is sufficiently larger than that of the term for maximizing the number of selected stocks. Furthermore, $\mu_\mathrm{R}$ was set to $2$. This value was chosen so that the expected return maximization term would have a certain influence on the optimization, allowing the model's performance to be clearly distinguished from that of the pure MIS model. The correlation threshold $\theta = 0.23$ was adopted as it was identified as an optimal value in the study by~\cite{hidaka2023correlation}. The return calculation period of $D_{\mathrm{R}} = 20$ was used to align with the monthly rebalancing strategy. The parameter $\alpha$, which is essential for calculating expected returns using EWAvg, was set to $0.003$ calculated from Eq.~\eqref{eq: alpha eqma}.
For solving the combinatorial optimization problems, we
employed a simulated-bifurcation-based solver~\cite{goto2019combinatorial, goto2021high}. We employed the solver's ballistic simulated bifurcation (bSB) algorithm and set the timeout to $10$ seconds.

\section{Results}
\label{sec: results}

\begin{figure}[t]
    \centering
    \includegraphics[width=1\linewidth]{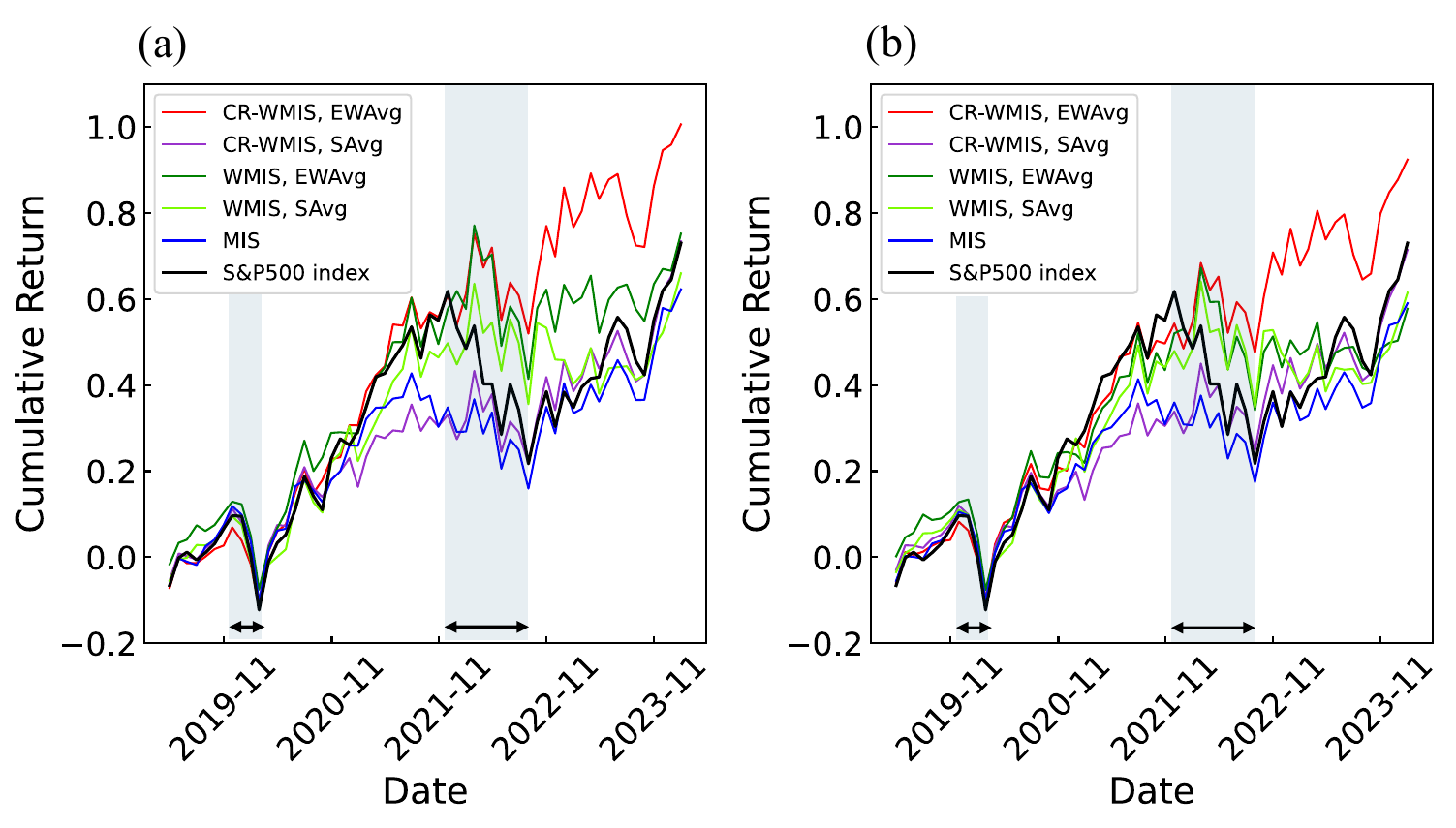}
    \caption{(Color online) Trajectory of cumulative return for each method over the five-year investment period (April 2019 to March 2024). Each line represents the average of 10 backtesting simulations. The shaded regions highlight the COVID-19 shock from February to April 2020 and the inflation and rising-interest-rate period from January to October 2022. Method-specific results with shaded areas representing standard deviation are provided in Appendix~\ref{appendix1: individual results}. (a) Equal-weight (EW) strategy, (b) inverse-volatility-weight (IVW) strategy.}
    \label{fig: cumret all}
\end{figure}

This section reports the results of the backtesting simulation performed based on the experimental setup described in the previous section.
First, we evaluate the performance of each method over the entire validation period from April 2019 to March 2024.
Figure~\ref{fig: cumret all} illustrates the cumulative return trajectories for each portfolio strategy and the market index (S\&P 500).
To maintain the readability of Fig.~\ref{fig: cumret all}, the variability across the backtesting simulations is not shown in the main figure. The corresponding cumulative return trajectories with shaded areas representing one standard deviation for each method under the EW allocation strategy are provided in Fig.~\ref{fig: individual results} in Appendix~\ref{appendix1: individual results}.
As shown in Fig.~\ref{fig: cumret all}, during the five-year investment period, the CR-WMIS model using EWAvg for expected return estimation achieved the highest final cumulative return, demonstrating superior profitability. A closer examination of the cumulative return trajectories reveals that the superiority of CR-WMIS is evident in both return and risk aspects. 
First, with respect to risk aversion, the CR-WMIS with EWAvg exhibited significantly lower drawdowns than the market index and the WMIS model during market downturns, such as the COVID-19 shock in 2020 and the period of inflation and rising interest rates in 2022. This suggests that the model possesses downside risk tolerance. 
Because the differences among the methods during these market downturns are difficult to distinguish in the full five-year trajectories, enlarged views are provided in Appendix~\ref{app:market_downturns}.
Second, in terms of returns, it outperformed the MIS model during recovery phases and bull markets, while maintaining a growth rate comparable to those of the market index and the WMIS model. These results indicate that the CR-WMIS model exhibits the dual characteristic of mitigating losses during market downturns while efficiently capturing growth during upturns. Regarding the asset allocation strategies, the EW strategy yielded slightly better return performance than the IVW strategy.

\begin{figure}[t]
    \centering
    \includegraphics[width=0.95\linewidth]{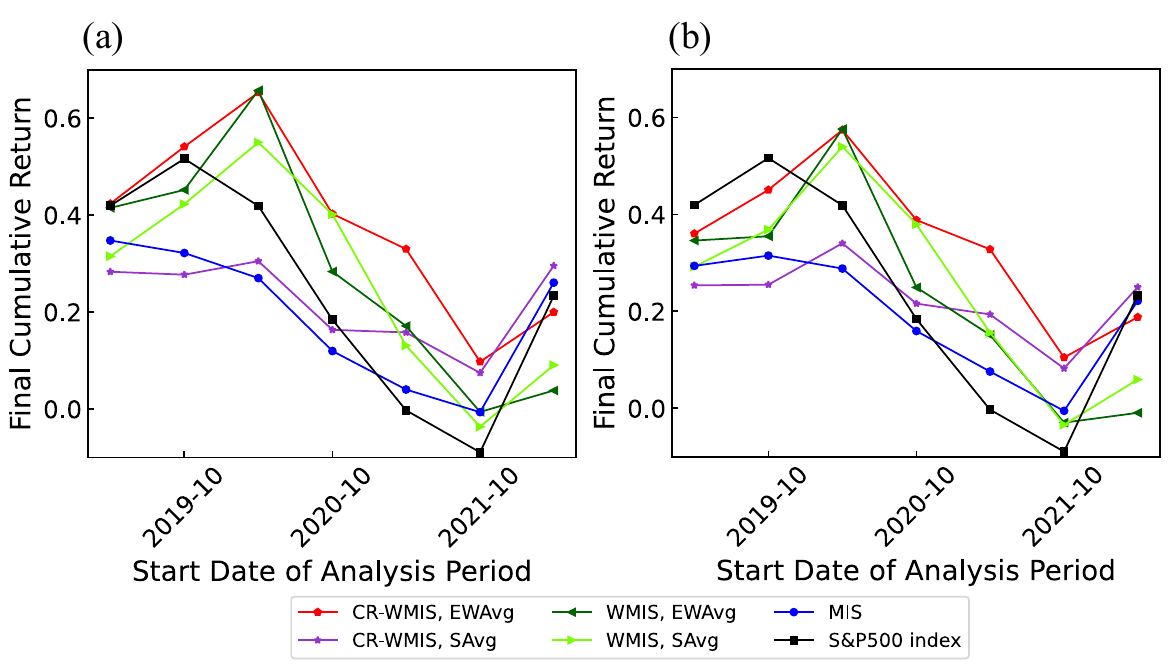}
    \caption{(Color online) Final cumulative return from the rolling-window analysis. Each data point is the average of 10 backtesting simulations for the corresponding time window. The horizontal axis indicates the start date of each window. (a) EW strategy, (b) IVW strategy.}
    \label{fig: final cumret}
\end{figure}

Next, we present the results of the rolling-window analysis, which was conducted to evaluate the robustness of our findings. First, we assess the robustness of the return characteristics. Figure~\ref{fig: final cumret} shows the final cumulative return for each two-year analysis period, where the starting time of the period is progressively advanced.
As shown in Fig.~\ref{fig: final cumret}, the CR-WMIS model using EWAvg for expected return estimation demonstrates superior return characteristics compared with the MIS and WMIS models, as well as the market index, in most of the time windows. These results indicate that the CR-WMIS model can generate stable and favorable returns across diverse market environments. Regarding the asset allocation strategies, as in the full-period analysis (Fig.~\ref{fig: cumret all}), the EW strategy again tended to outperform the IVW strategy from a returns perspective.

\begin{figure}[t]
    \centering
    \includegraphics[width=1\linewidth]{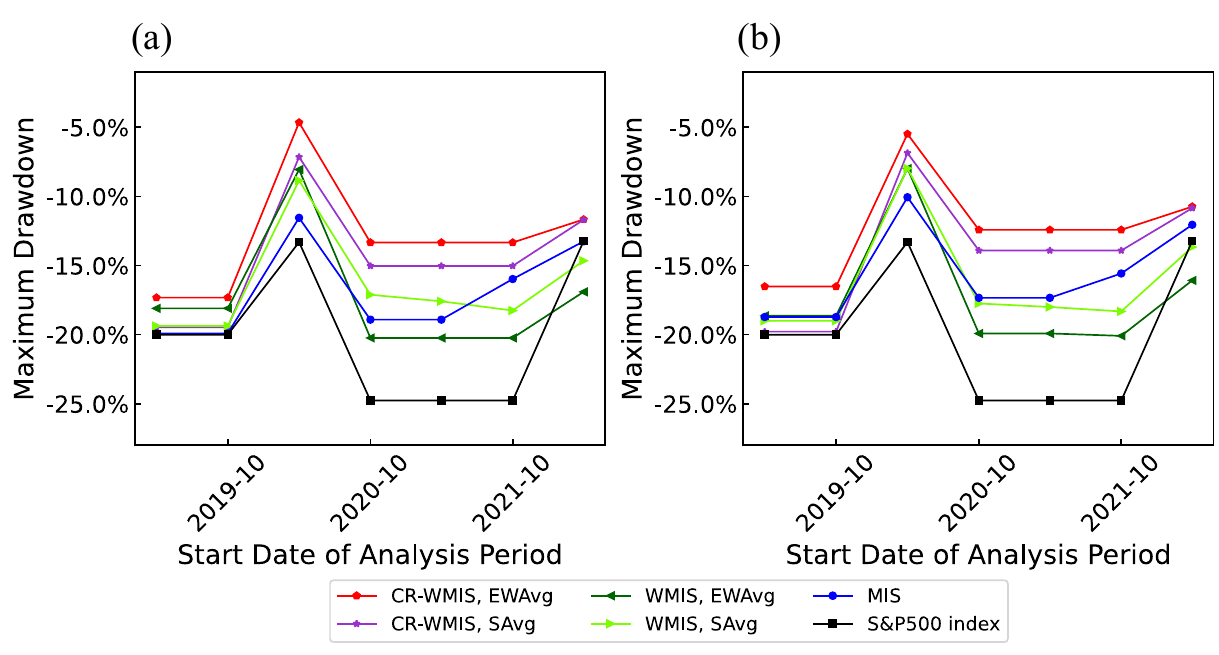}
    \caption{(Color online) Maximum drawdown (MDD) from the rolling-window analysis. Each data point is the average of 10 backtesting simulations for the corresponding time window. The horizontal axis indicates the start date of each window. (a) EW strategy, (b) IVW strategy.}
    \label{fig: mdd}
\end{figure}

\begin{figure}[t]
    \centering
    \includegraphics[width=1\linewidth]{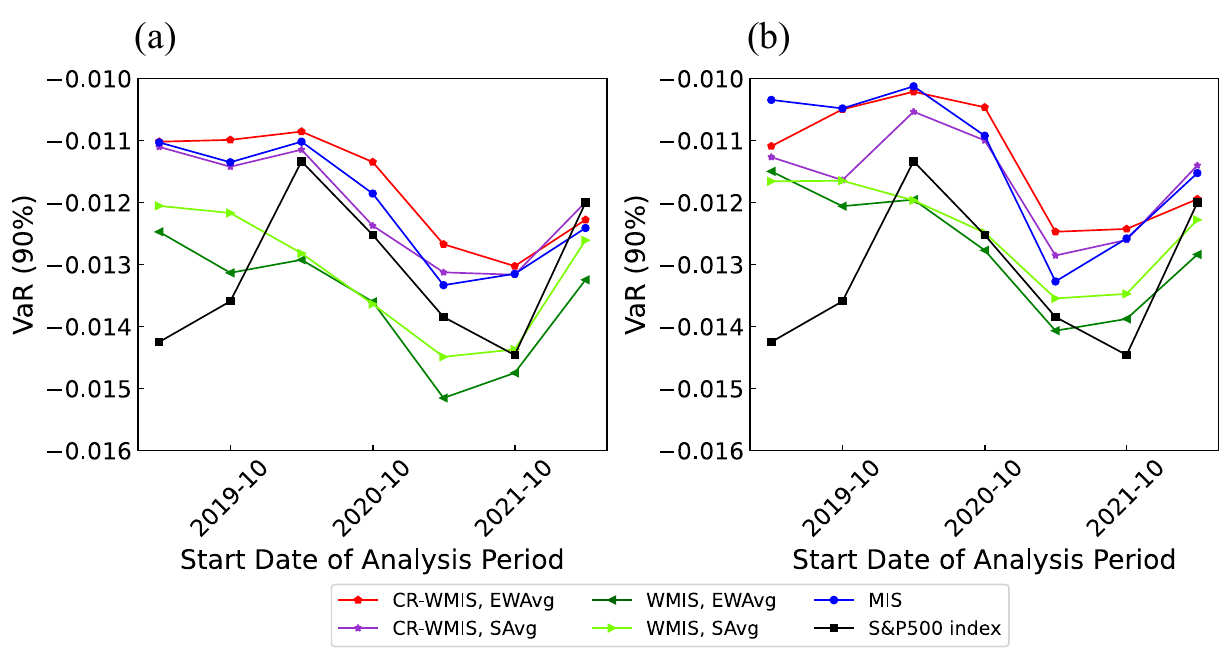}
    \caption{(Color online) Value at Risk, VaR ($q=0.1$) from the rolling-window analysis. Each data point is the average of 10 backtesting simulations for the corresponding time window. The horizontal axis indicates the start date of each window. (a) EW strategy, (b) IVW strategy.}
    \label{fig: var}
\end{figure}

\begin{figure}[t]
    \centering
    \includegraphics[width=1\linewidth]{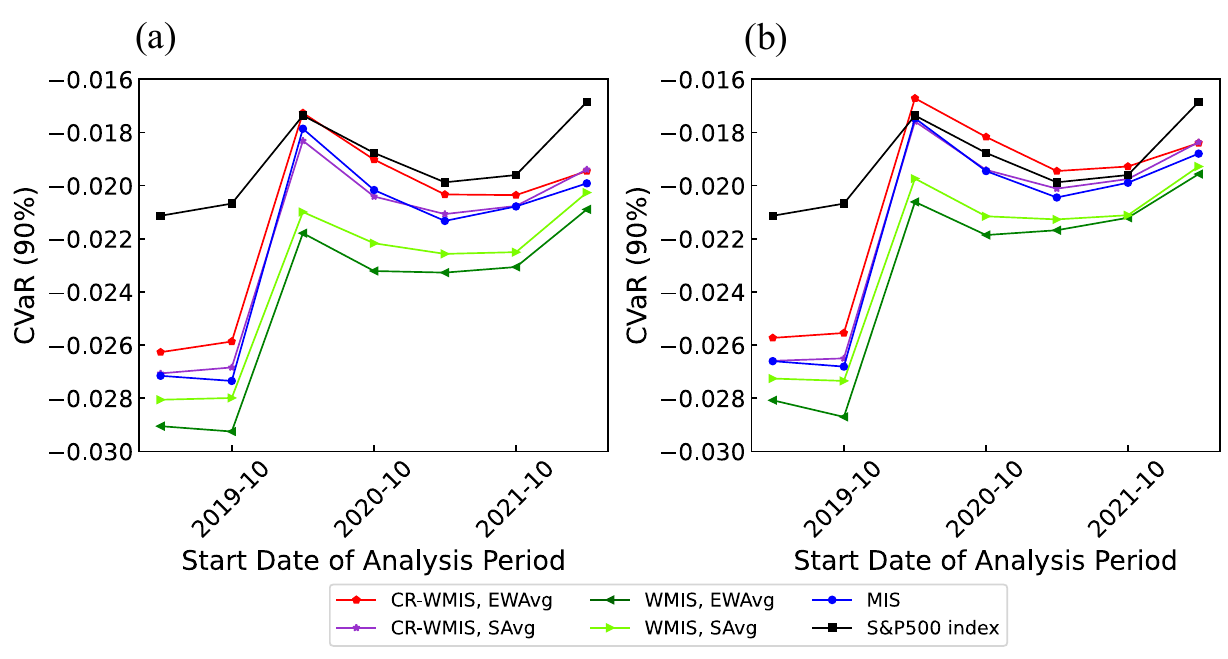}
    \caption{(Color online) Conditional Value at Risk, CVaR ($q=0.1$) from the rolling-window analysis. Each data point is the average of 10 backtesting simulations for the corresponding time window. The horizontal axis indicates the start date of each window. (a) EW strategy, (b) IVW strategy.}
    \label{fig: cvar}
\end{figure}

\begin{figure}[t]
    \centering
    \includegraphics[width=0.97\linewidth]{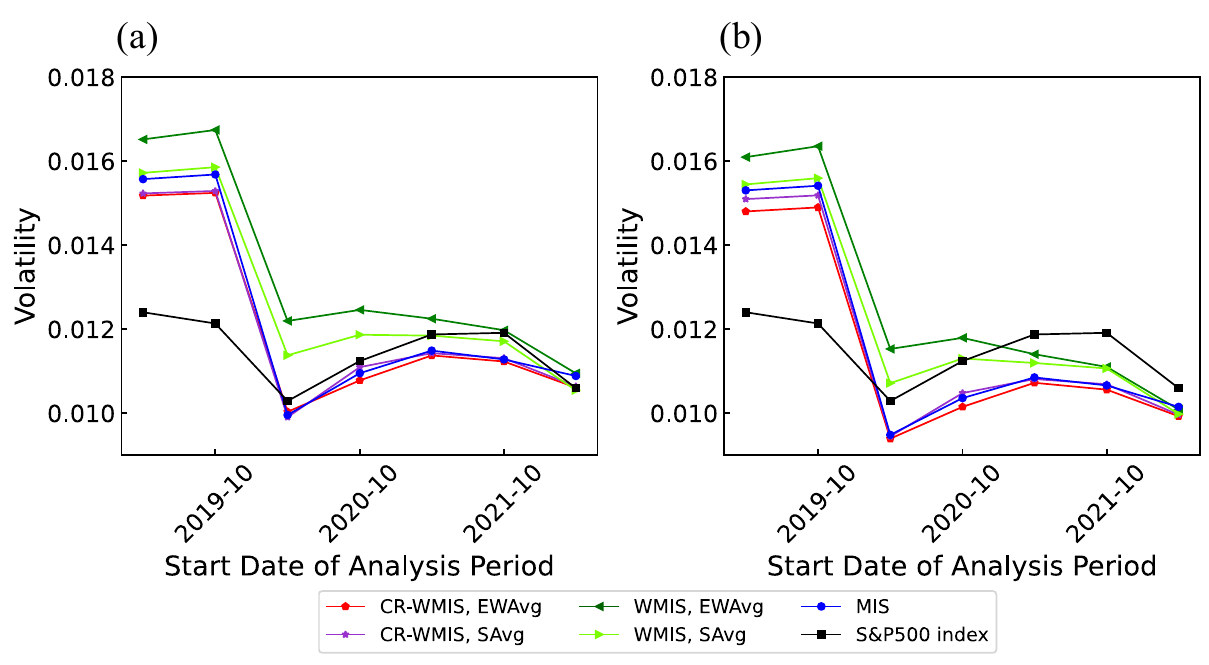}
    \caption{(Color online) Volatility from the rolling-window analysis. Each data point is the average of 10 backtesting simulations for the corresponding time window. The horizontal axis indicates the start date of each window. Note that a lower volatility value is considered more favorable. (a) EW strategy, (b) IVW strategy.}
    \label{fig: volatility}
\end{figure}

Subsequently, we assess the robustness of the risk characteristics. Figs.~\ref{fig: mdd}-\ref{fig: volatility} present comparisons of MDD, VaR, CVaR, and volatility across each time window.
According to Figs.~\ref{fig: mdd}, \ref{fig: var}, and \ref{fig: volatility}, for the three metrics of MDD, VaR, and volatility, the CR-WMIS model exhibits superior risk characteristics compared to the other methods and the market index in most time windows. On the other hand, as shown in Fig.~\ref{fig: cvar}, although the CVaR of the CR-WMIS model was occasionally inferior to that of the market index, it consistently outperformed the MIS and WMIS models. These results indicate that CR-WMIS consistently exhibits strong risk-aversion effects across multiple risk metrics. From an asset allocation perspective, in contrast to the return characteristics, the IVW strategy generally achieved better risk performance than the EW strategy. This finding indicates that the IVW strategy effectively mitigates overall portfolio risk by adjusting investment weights according to each stock's volatility, thereby allocating more capital to lower-risk assets.

Another noteworthy observation is that the MIS model tended to exhibit better risk characteristics than the WMIS model. This finding suggests that maximizing the number of selected stocks, the core principle of the MIS strategy, contributes to effective risk diversification. Therefore, CR-WMIS can therefore be regarded as an approach that inherits this intrinsic risk-aversion capability from MIS while simultaneously achieving enhanced returns.

From the above analysis, it is evident that the CR-WMIS demonstrates superior performance for both return and risk aspects. This advantage suggests that the weighting by expected return not only led to improved returns but also preserved the inherent risk-aversion effect of the MIS approach. The superiority of CR-WMIS is discussed in more detail in Section~\ref{sec: discussion}.

\section{Discussion}
\label{sec: discussion}

As shown in Section~\ref{sec: results}, our proposed CR-WMIS method demonstrated superior return and risk characteristics, outperforming the existing MIS and WMIS models, as well as the market index (S\&P 500), in a five-year backtesting simulation. Building on these findings, this section provides a multifaceted analysis of the factors underlying to the superiority of CR-WMIS and elucidates the mechanisms behind its performance improvement.

First, in Section~\ref{subsec: discussion1}, we investigate the drivers of the model's performance by examining the relationship between the characteristics of its in-sample solutions (derived from historical data used for optimization) and its actual out-of-sample performance (on unseen future data). Next, in Section~\ref{subsec: discussion2}, we conduct a sensitivity analysis of the key model parameters to quantitatively evaluate the trade-off between risk diversification through cardinality maximization and profitability improvement through expected return maximization.

\subsection{Mechanism of Performance Improvement}
\label{subsec: discussion1}

To elucidate the factors underlying the performance improvement, a two-step validation process is essential. The first step involves evaluating the in-sample solutions. We verify whether the solver effectively identifies solutions with high expected returns, in accordance with the objective function constructed from historical (in-sample) data. The second step assesses the persistence of this performance using out-of-sample data. We examine whether the portfolio identified as optimal in the in-sample data also exhibits strong performance on unseen, future (out-of-sample) data. This allows evaluation of whether the in-sample expected return serves as an effective predictor of future performance.

\begin{figure}
    \centering
    \includegraphics[width=1.0\linewidth]{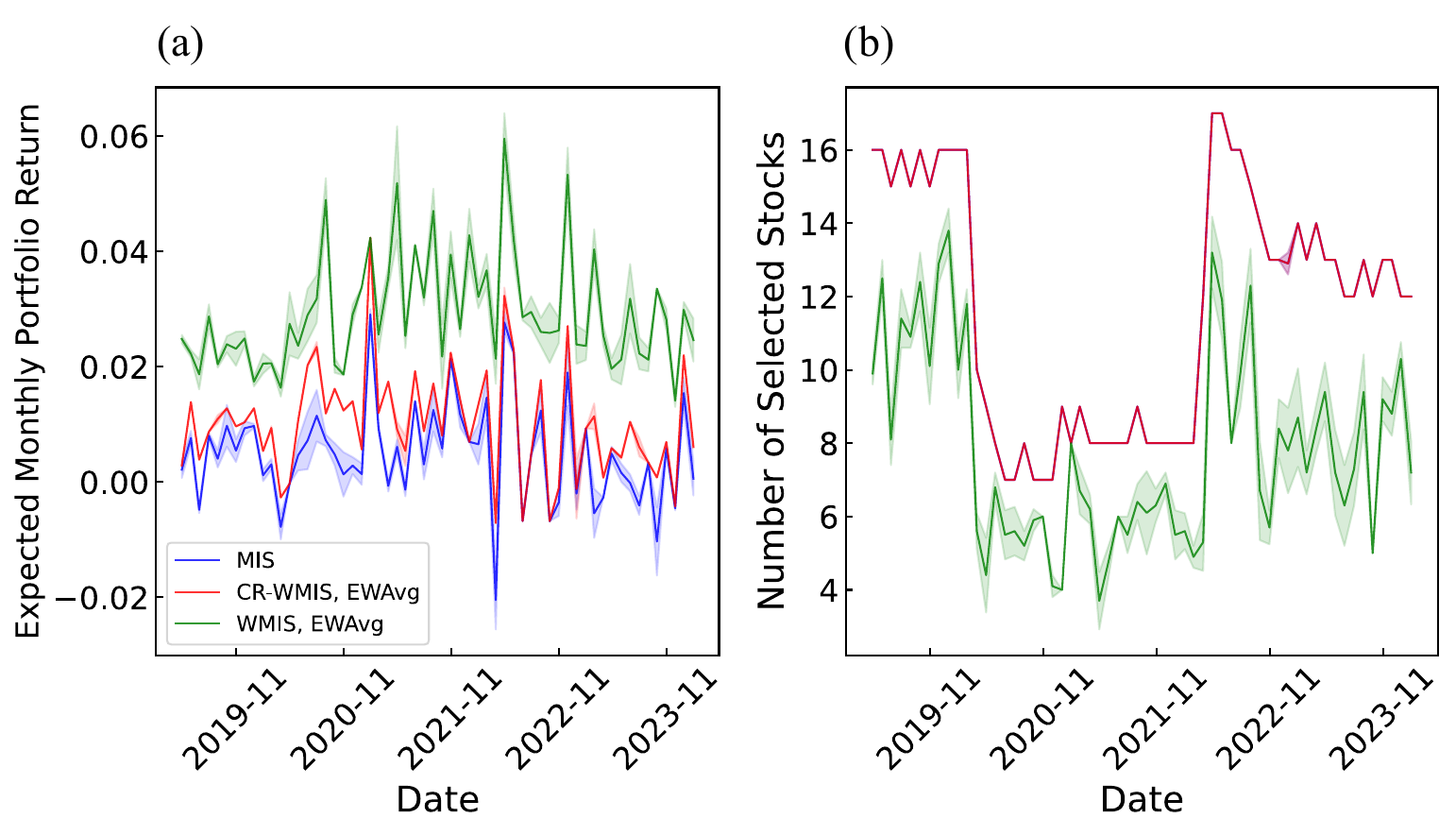}
    \caption{(Color online) Evaluation of in-sample solution quality for each method. The solid lines represent the mean, and the shaded areas indicate the standard deviation from 10 backtesting simulations. (a) Trajectory of the portfolio's in-sample expected return, which is calculated with EW, (b) Trajectory of the number of selected stocks. Note: The lines for MIS (blue) and CR-WMIS (red) in panel (b) overlap because their values for the number of selected stocks are identical.}
    \label{fig: expected return and num stocks}
\end{figure}

First, we analyze the characteristics of the solutions that each method generates from the in-sample data. Figure~\ref{fig: expected return and num stocks}(a) shows the trajectory of the portfolio's expected return, defined as the average of the expected returns of the selected stocks, $\sum_{p \in S_{\mathrm{selected}}} r_p / N_\mathrm{portfolio}$, which is calculated from the set of selected stocks $S_{\mathrm{selected}} = \{i \mid x_i^*=1\}$. Figure~\ref{fig: expected return and num stocks}(b) presents the trajectory of the number of selected stocks $N_\mathrm{portfolio} = |S_{\mathrm{selected}}|$. We have confirmed that all solutions are valid independent sets; that is, they satisfy the condition $\sum_{1 \leq i < j \leq N} f_{i,j} x_i^* x_j^* = 0$.
As shown in Fig.~\ref{fig: expected return and num stocks}(a), CR-WMIS consistently achieves higher in-sample expected returns than MIS throughout the entire investment period. Furthermore, Fig.~\ref{fig: expected return and num stocks}(b) indicates that the number of selected stocks in CR-WMIS and MIS is identical. This result demonstrates that CR-WMIS retains the core principle of MIS, which is the risk diversification effect from maximizing the number of selected stocks, while simultaneously enhancing profitability through expected-return-based weighting.

In contrast, a comparison with WMIS reveals contrasting characteristics. As shown in Fig.~\ref{fig: expected return and num stocks}(a), WMIS exhibits a significantly higher in-sample expected return than CR-WMIS. However, Fig.~\ref{fig: expected return and num stocks}(b) indicates that the number of selected stocks for WMIS is substantially reduced. This finding implies that WMIS, in its exclusive pursuit of maximizing expected return, sacrifices the diversification effect that lies at the core of the MIS strategy. This reduction in the number of selected stocks is considered the primary factor contributing to the deterioration of the WMIS model's risk characteristics, as suggested in Section~\ref{sec: results}.

\begin{figure}
    \centering
    \includegraphics[width=1\linewidth]{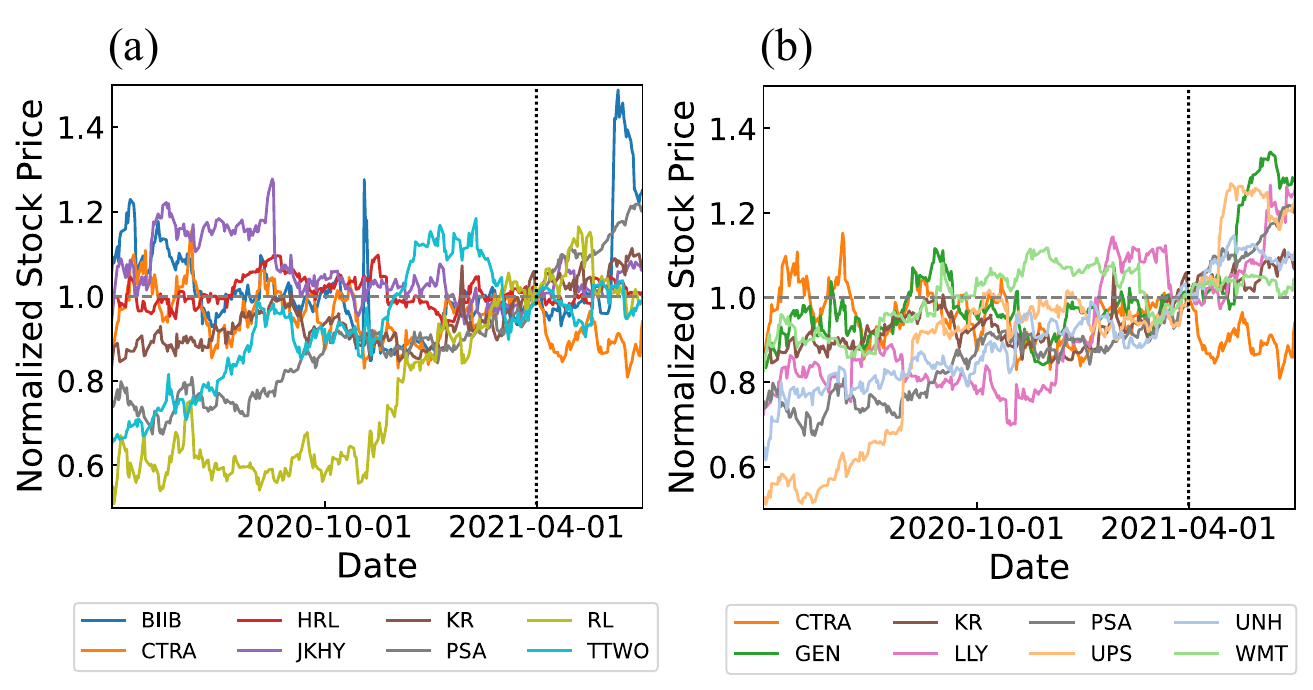}
    \caption{(Color online) Trajectories of stock prices for selected stocks, with the optimization date of April 1, 2021,indicated by the dotted black line. The vertical axis shows the stock price normalized by its value on the optimization date, allowing for a comparison of growth rate. The legend entries are the ticker symbols of the S\&P 500 stocks. (a) Stocks selected by MIS. (b) Stocks selected by CR-WMIS with EWAvg.}
    \label{fig: stock price}
\end{figure}

Next, we examine the relationship between the optimization results obtained from in-sample data and the subsequent performance on the out-of-sample data. Figure~\ref{fig: stock price} presents the stock price trajectories of the stocks selected by the MIS and CR-WMIS models. For this illustration, we use a representative optimization date of April 1, 2021, the midpoint of our investment period. To focus on the price growth rate, all stock prices are normalized to their value on the optimization date.
As shown in Fig.~\ref{fig: stock price}, CR-WMIS, compared to MIS, selects a greater number of stocks that exhibited an upward trend during the in-sample period (prior to the optimization date). This indicates that the weighting by expected return functions effectively. Furthermore, many of the stocks that exhibited upward trend in the in-sample period continued to show similar behavior into the out-of-sample period. This suggests the existence of persistence, or a ``momentum effect'' in stock price trends at the monthly time scale of our rebalancing strategy.

Since this result is an example from a single point in time, it does not guarantee that the in-sample and out-of-sample trends will always align. However, two comprehensive pieces of evidence support the generality of this consistency. First, Fig.~\ref{fig: expected return and num stocks}(a) demonstrates that CR-WMIS maintains high in-sample expected returns throughout the investment period. Second, as shown in Figs.~\ref{fig: cumret all} and \ref{fig: final cumret}, CR-WMIS consistently exhibits superior performance in terms of cumulative return in the long-term simulations. Taken together, these results indicate that the superiority of CR-WMIS can be explained by the following mechanism. With respect to the enhancement of return characteristics, CR-WMIS preferentially selects stocks exhibiting high growth in the in-sample data as a result of its weighting based on expected return. Consequently, because stocks with high in-sample growth tend to maintain their growth in the out-of-sample period on a monthly rebalancing time scale, CR-WMIS achieves superior out-of-sample return performance.

Next, with respect to the risk characteristics, CR-WMIS preserves the effect of maximizing the number of selected stocks, a feature inherited from MIS, even while introducing weighting based on expected return. This enables the model to retain a diversification effect absent in the WMIS model, thereby achieving strong risk-aversion capabilities. As shown in Fig.~\ref{fig: expected return and num stocks}(b), although the number of selected stocks decreases significantly for WMIS, CR-WMIS maintains a number of stocks comparable to that of MIS. This difference is considered the primary cause of the disparity in their risk characteristics. Although WMIS controls risk based on inter-stock correlations via the independent set constraint, it does not account for the number of stocks, making it prone to concentrated investments in a few high-return stocks. In contrast, because CR-WMIS incorporates the maximization of the number of selected stocks as an additional objective alongside the independent set constraint, it achieves broader diversification and enhanced robustness against idiosyncratic stock risk and parameter estimation errors.

Through this integrated performance enhancement mechanism, CR-WMIS integrates the strengths of conventional methods while overcoming their limitations. Therefore, it can be concluded that CR-WMIS demonstrates superiority in both return and risk aspects.

\subsection{Analysis of Parameter Sensitivity}
\label{subsec: discussion2}

\begin{figure}
    \centering
    \includegraphics[width=1\linewidth]{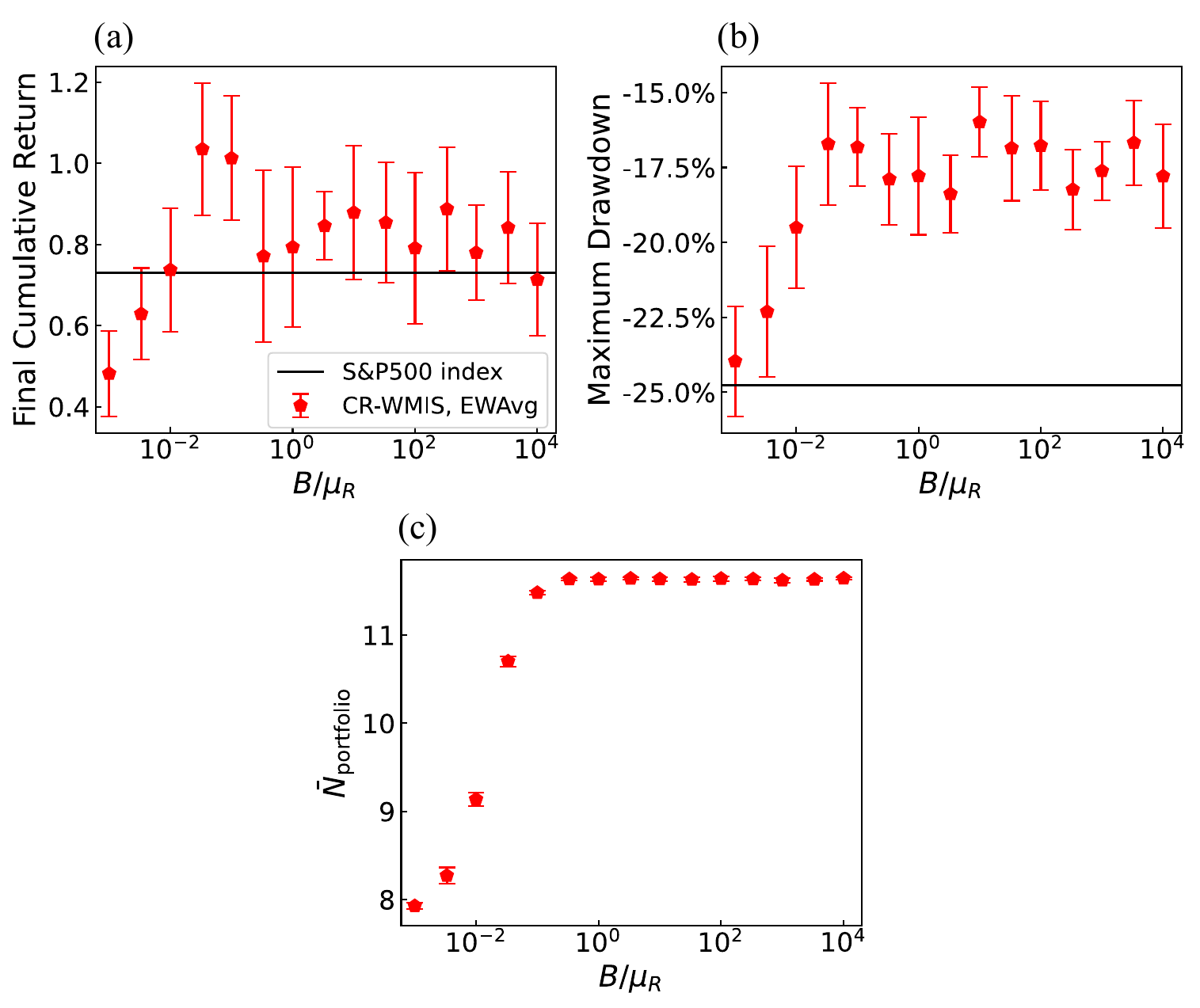}
    \caption{(Color online) Dependence of performance metrics on the ratio $B/\mu_\mathrm{R}$ for the EW strategy. Data points and error bars represent the mean and standard deviation, respectively, from 10 backtesting simulations. The black horizontal line indicates the S\&P 500 index benchmark. The horizontal axis is on a logarithmic scale. (a) Final cumulative return, (b) maximum drawdown, (c) average number of selected stocks over the entire backtesting period, $\bar{N}_\mathrm{portfolio}$.}
    \label{fig: B by muR dependence}
\end{figure}

In this section, we analyze the dependence of the proposed CR-WMIS model’s performance on its key parameters. The CR-WMIS model, defined in Eq.~\eqref{eq: cr-wmis qubo}, consists of three components: the independent set constraint term, the cardinality maximization term, and the expected return maximization term. Since the independent set constraint is fundamental to the MIS, WMIS, and CR-WMIS approaches, we set the coefficient $A$ to be sufficiently larger than the other coefficients, $B$ and $\mu_\mathrm{R} r_i$, to ensure this constraint is always satisfied. On the other hand, the relative magnitude of $B$ and $\mu_\mathrm{R} r_i$ provides flexibility in adjusting the trade-off between risk diversification via cardinality maximization and profitability improvement through expected return maximization. Therefore, to identify an appropriate balance, this section investigates how performance metrics depend on the ratio $B / \mu_\mathrm{R}$ in the backtesting simulation. Figure~\ref{fig: B by muR dependence} shows the results for (a) final cumulative return, (b) maximum drawdown, and (c) the average number of selected stocks over the entire backtesting period, $\bar{N}_\mathrm{portfolio}$.
Figure~\ref{fig: B by muR dependence}(a) presents that the final cumulative return decreases when the ratio $B / \mu_\mathrm{R}$ is either extremely large or small. Furthermore, Fig.~\ref{fig: B by muR dependence}(b) reveals a trend of worsening maximum drawdown in the region where $B / \mu_\mathrm{R}$ is small. This behavior can be attributed to the balance among the terms in the model's objective function.

First, when $B \ll \mu_\mathrm{R}$, the CR-WMIS model asymptotically converges to the WMIS model, which prioritizes expected return maximization. As shown in Fig.~\ref{fig: B by muR dependence}(c), as a result, investments become concentrated in a few high-return stocks, and the number of selected stocks decreases. This results in insufficient portfolio diversification and consequently, despite a high in-sample expected return, the actual risk increases, which likely resulted in the large drawdowns observed.

Conversely, when $B \gg \mu_\mathrm{R}$, the CR-WMIS model converges toward the MIS model, which aims to maximize cardinality. In this setting, as shown in Fig.~\ref{fig: B by muR dependence}(c), the model selects the maximum possible number of stocks permitted by the independent set constraint. Although this achieves strong diversification and suppresses risk, expected return is not prioritized, leading to missed profit opportunities and, ultimately, a lower final cumulative return.

From these results, it is evident that the best performance in both return and risk is achieved within a parameter region where $B / \mu_\mathrm{R}$ is neither too large nor too small. Specifically, Figs.~\ref{fig: B by muR dependence}(a) and (b) indicate that the optimal region lies around $B / \mu_\mathrm{R} \approx 10^{-1}$. As shown in Fig.~\ref{fig: B by muR dependence}(c), this parameter region is located at the boundary between the regime where the number of selected stocks is maintained at its maximum and the zone where it begins to decrease in favor of higher expected returns. This finding suggests that this point represents the optimal balance between risk diversification through cardinality maximization and profitability improvement through expected return maximization. This optimal ratio of $B / \mu_\mathrm{R} \approx 10^{-1}$ is likely associated with the fact that the maximum value of the expected returns $r_i$ in the dataset is on the order of $10^{-1}$. This implies that, to maximize the performance of the CR-WMIS model, the cardinality coefficient $B$ should be set to the same order of magnitude as the maximum value of the expected return term, $\max_i(\mu_\mathrm{R} r_i)$. 

\section{Conclusion}
\label{sec: conclusion}

In this study, we proposed the cardinality-return weighted maximum independent set (CR-WMIS) approach for the financial portfolio optimization problem and conducted a comprehensive validation of its effectiveness. The main contributions of this research are summarized as follows. First, we proposed the new CR-WMIS model that integrates the conventional MIS and WMIS models. This model optimizes both return and risk by integrating the diversification effect from maximizing the number of selected stocks with the weighting effect based on expected returns. Second, we conducted a comprehensive risk assessment of graph-theoretic portfolio optimization method, an area that has been insufficiently explored in prior research. Specifically, by using multiple risk metrics including maximum drawdown, VaR, CVaR, and volatility, we performed a multifaceted validation of not only its return characteristics but also its risk-aversion capabilities. The results indicated the effectiveness of CR-WMIS in terms of both return and risk, providing valuable insights for its practical application. Third, we confirmed the effectiveness of the exponentially weighted average (EWAvg) for estimating expected returns within the CR-WMIS formulation, achieving superior performance compared to the simple average (SAvg). Fourth, we examined two asset allocation methods for CR-WMIS, equal-weight (EW) and inverse-volatility-weight (IVW), and found that the EW strategy was advantageous for returns, whereas the IVW strategy was superior for managing risk.

The findings of this study provide important implications for portfolio optimization theory. The conventional Markowitz mean-variance model has fundamental limitations, including its assumption of normally distributed returns and its high sensitivity to estimation errors in input parameters. These factors reduce its practical applicability, given the non-normality and presence of extreme values in actual financial market return distributions. In contrast, graph-theoretic approaches such as MIS and our proposed CR-WMIS provide a fundamentally different framework for addressing these problems, employing a graph-theoretic independent set constraint and maximizing the number of selected stocks. In particular, maximizing the number of selected stocks mitigates the impact of extreme price movements of individual stocks and the effect of estimation errors through diversification. The comprehensive risk assessment conducted in this study yields a key insight not clarified in previous research: we demonstrated through multiple risk metrics that the maximization of selected stocks, a core concept from MIS, is not merely theoretical but produces a measurable risk-reduction effect in real market environments. This finding indicates that graph-theoretic methods are effective for practical investment decision-making in real-world markets. The results of the monthly rebalancing simulation showed that CR-WMIS maintained consistent superiority throughout the investment period, even against market shocks of various types, such as the COVID-19 pandemic and the period of high inflation and rising interest rates. These findings demonstrate that our proposed method is a practical portfolio optimization framework that can deliver stable performance across diverse market conditions.

Future research should focus on validating the robustness of the method under different rebalancing frequencies and across various markets. Furthermore, the return-risk profile could be further enhanced by employing machine learning or Monte Carlo simulation techniques to estimate expected returns and correlations. Finally, comparing the proposed approach with more recent and advanced extensions of the mean–variance model represents an important direction for future research.

\section*{acknowledgement}
This work was partially supported by the Council for Science, Technology, and Innovation (CSTI) through the Cross-ministerial Strategic Innovation Promotion Program (SIP), ``Promoting the application of advanced quantum technology platforms to social issues'' (Funding agency: QST), the Japan Society for the Promotion of Science (JSPS) KAKENHI (Grant Number JP23H05447), Japan Science and Technology Agency (JST) (Grant Number JPMJPF2221). S. Tanaka wishes to express their gratitude to the World Premier International Research Center Initiative (WPI), MEXT, Japan, for their support of the Human Biology-Microbiome-Quantum Research Center (Bio2Q).

\bibliographystyle{jpsj}
\bibliography{reference}
\appendix
\section{Variability of Cumulative Return Trajectories}
\label{appendix1: individual results}
\begin{figure}
    \centering
    \includegraphics[width=1\linewidth]{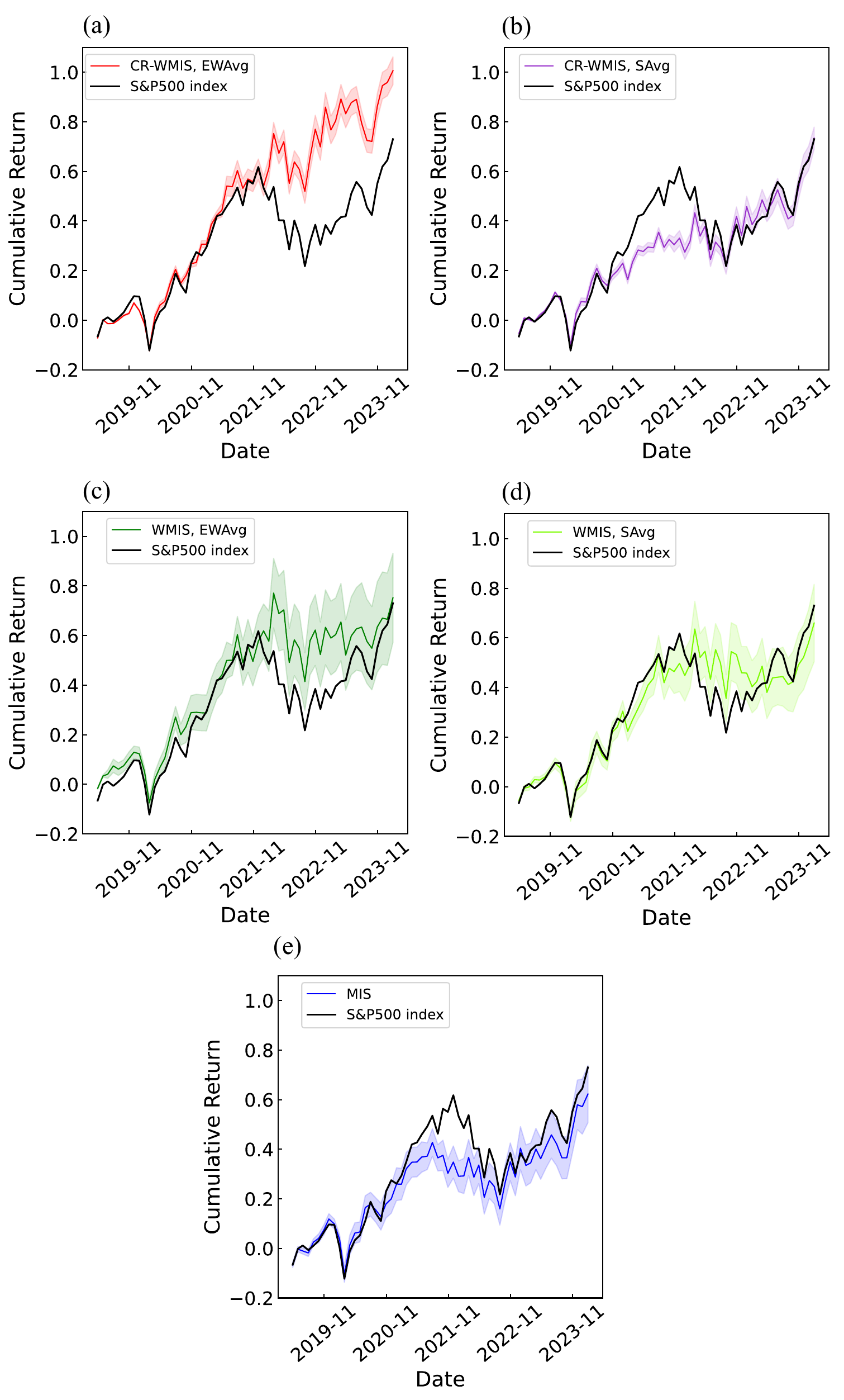}
    \caption{(Color online) 
    Cumulative return trajectories and their variability under the equal-weight (EW) allocation strategy. The solid lines represent the mean cumulative returns over 10 backtesting simulations, and the shaded areas represent standard deviation. The S\&P 500 index is shown as a market benchmark in each panel.
    (a) CR-WMIS with EWAvg,
    (b) CR-WMIS with SAvg,
    (c) WMIS with EWAvg,
    (d) WMIS with SAvg, and
    (e) MIS.}
    \label{fig: individual results}
\end{figure}
To provide additional information on the variability of the backtesting results, Fig.~\ref{fig: individual results} presents the cumulative return trajectories for each optimization method under the equal-weight (EW) allocation strategy. The solid lines represent the mean cumulative returns obtained from 10 backtesting simulations, and the shaded areas represent standard deviation across these simulations. The trajectory of the S\&P 500 index is also included in each panel as a common market benchmark.
The results are presented separately for (a) CR-WMIS with EWAvg, (b) CR-WMIS with SAvg, (c) WMIS with EWAvg, (d) WMIS with SAvg, and (e) MIS. Presenting the results separately provides a detailed view of the variability associated with each method while avoiding excessive visual complexity in Fig.~\ref{fig: cumret all}.

\section{Detailed Performance during Major Market Downturns}
\label{app:market_downturns}

\begin{figure}
    \centering
    \includegraphics[width=1\linewidth]{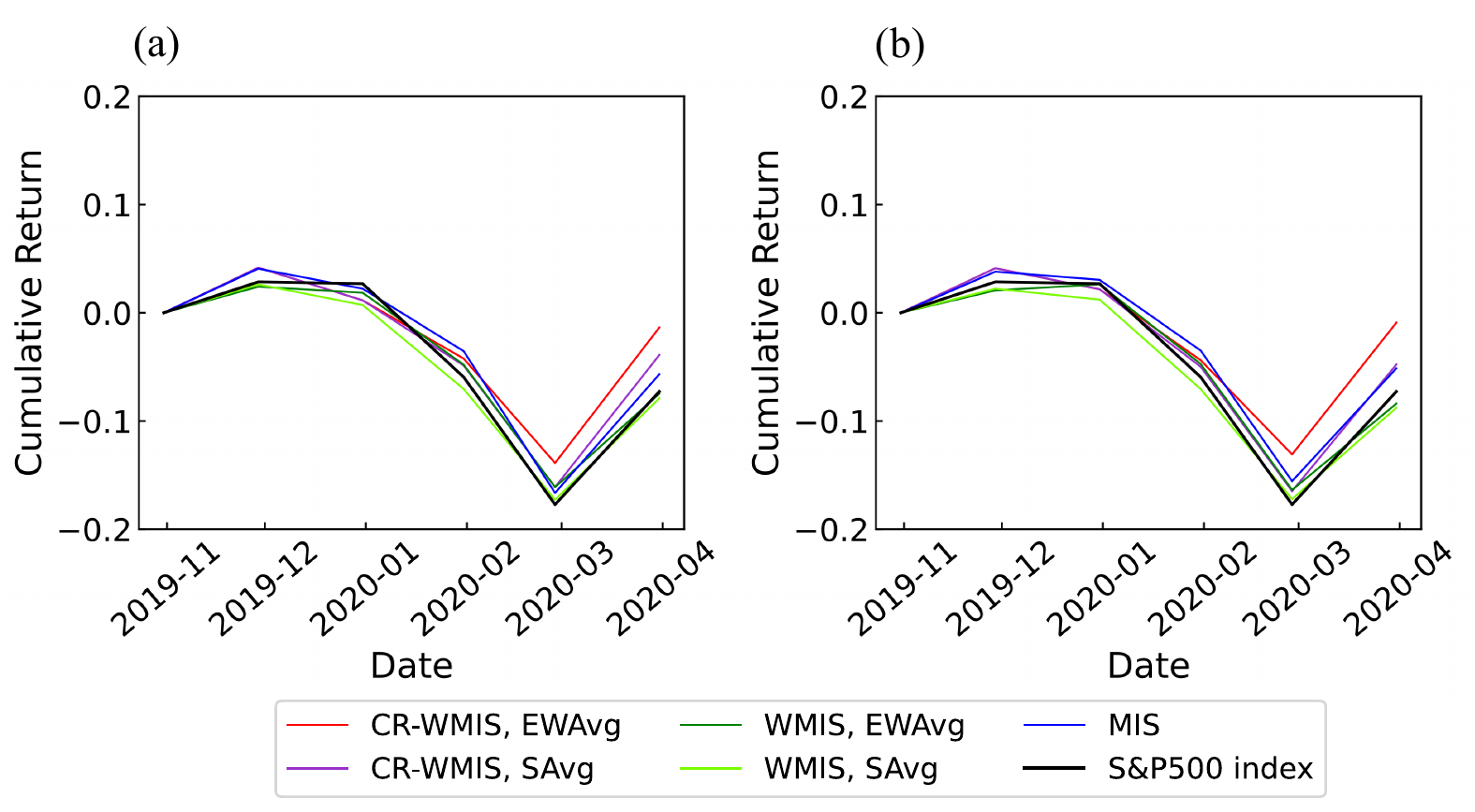}
    \caption{(Color online) 
    Enlarged cumulative return trajectories around the COVID-19 shock. Each line represents the mean of 10 backtesting simulations. The same five optimization methods as those shown in Fig.~\ref{fig: cumret all}, together with the S\&P 500 index, are presented. 
    (a) EW strategy, (b) IVW strategy.}
    \label{fig:covid19_zoom}
\end{figure}

\begin{figure}
    \centering
    \includegraphics[width=1\linewidth]{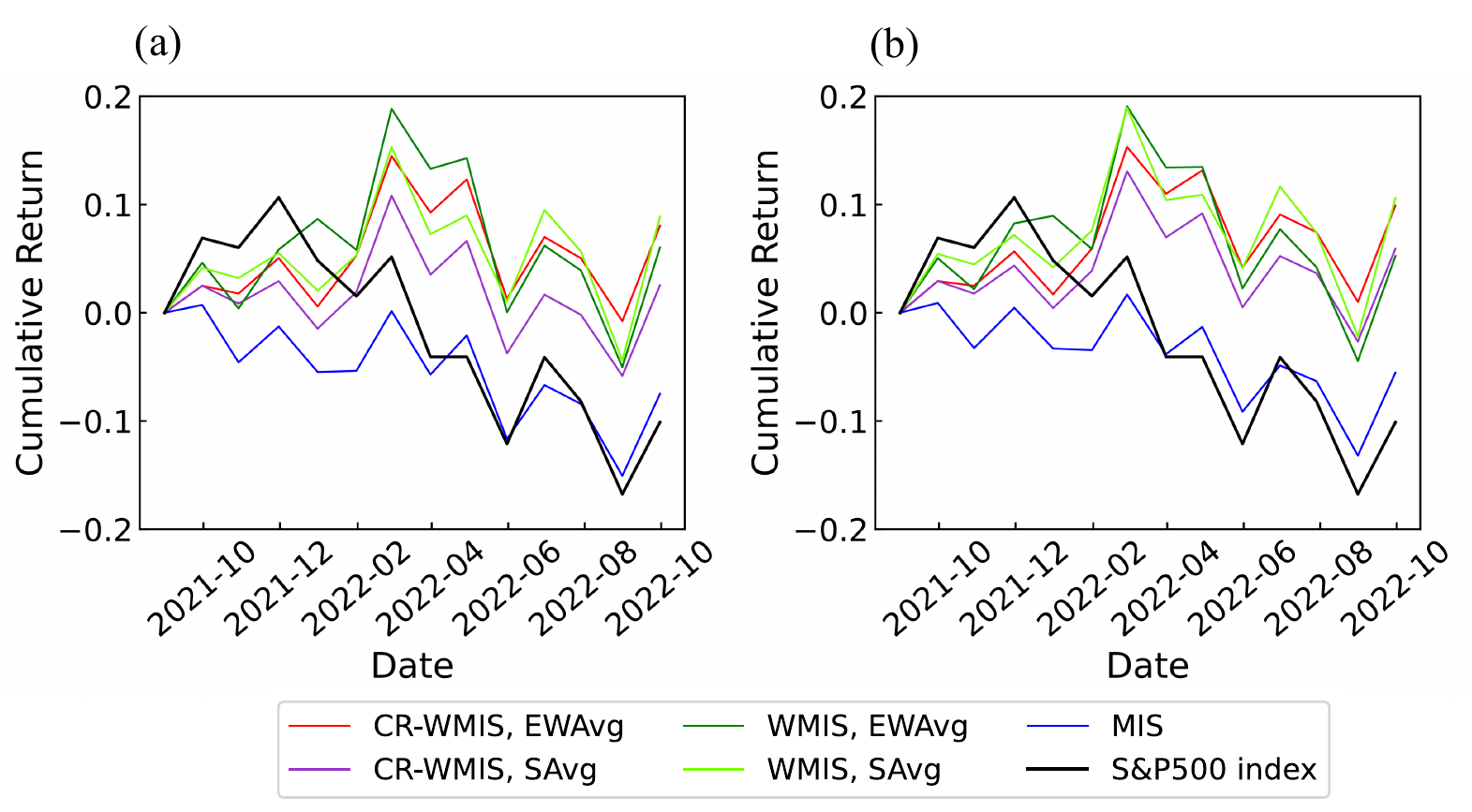}
    \caption{(Color online) 
    Enlarged cumulative return trajectories around the inflation and rising-interest-rate period. Each line represents the mean of 10 backtesting simulations. The same five optimization methods as those shown in Fig.~\ref{fig: cumret all}, together with the S\&P 500 index, are presented.
    (a) EW strategy, (b) IVW strategy.}
    \label{fig:inflation_zoom}
\end{figure}

To facilitate a more detailed comparison of the performance of the portfolio optimization methods during major market downturns, Figs.~\ref{fig:covid19_zoom} and~\ref{fig:inflation_zoom} present enlarged cumulative return trajectories around the COVID-19 shock and the inflation and rising-interest-rate period, respectively. Each figure shows the results for both the equal-weight (EW) and inverse-volatility-weight (IVW) allocation strategies. The same five optimization methods as those shown in Fig.~\ref{fig: cumret all}, together with the S\&P 500 index, are included for comparison.
Figure~\ref{fig:covid19_zoom} shows the cumulative return trajectories from November 2019 to April 2020, including the sharp market decline associated with the COVID-19 shock. Under both allocation strategies, CR-WMIS with EWAvg experienced the smallest drawdown among the compared methods and the market index. This enlarged view more clearly demonstrates the downside-risk mitigation achieved by CR-WMIS with EWAvg during the COVID-19 shock.
Figure~\ref{fig:inflation_zoom} shows the cumulative return trajectories from September 2021 to October 2022, covering the market downturn associated with inflation and rising interest rates. During this period, the two CR-WMIS models and the two WMIS models experienced relatively limited drawdowns, whereas the MIS model and the S\&P 500 index exhibited substantially larger declines. These results indicate that expected-return-based weighting contributed to mitigating losses during this market environment.


\end{document}